\documentclass[preprint2]{aastex631}

\usepackage{comment}

\begin{document}

\title{JWST-TST DREAMS: NIRSpec/PRISM Transmission Spectroscopy of the Habitable Zone Planet TRAPPIST-1 e}

\author[0000-0001-9513-1449]{N\'estor Espinoza}
\affiliation{Space Telescope Science Institute, 3700 San Martin Drive, Baltimore, MD 21218, USA}
\affiliation{William H. Miller III Department of Physics and Astronomy, Johns Hopkins University, Baltimore, MD 21218, USA}

\author[0000-0002-0832-710X]{Natalie H. Allen}
\affiliation{William H. Miller III Department of Physics and Astronomy, Johns Hopkins University, Baltimore, MD 21218, USA}

\author[0000-0002-5322-2315]{Ana Glidden}
\affiliation{Department of Earth, Atmospheric and Planetary Sciences, Massachusetts Institute of Technology, Cambridge, MA 02139, USA}
\affiliation{Kavli Institute for Astrophysics and Space Research, Massachusetts Institute of Technology, Cambridge, MA 02139, USA} 

\author[0000-0002-8507-1304]{Nikole K. Lewis}
\affiliation{Department of Astronomy and Carl Sagan Institute, Cornell University, 122 Sciences Drive, Ithaca, NY 14853, USA}

\author[0000-0002-6892-6948]{Sara Seager}
\affiliation{Department of Earth, Atmospheric and Planetary Sciences, Massachusetts Institute of Technology, Cambridge, MA 02139, USA}
\affiliation{Kavli Institute for Astrophysics and Space Research, Massachusetts Institute of Technology, Cambridge, MA 02139, USA} 
\affiliation{Department of Aeronautics and Astronautics, MIT, 77 Massachusetts Avenue, Cambridge, MA 02139, USA}


\author[0000-0003-4835-0619]{Caleb I. Ca\~{n}as}
\affiliation{NASA Goddard Space Flight Center, Greenbelt, MD 20771, USA}

\author[0000-0001-5878-618X]{David Grant}
\affiliation{University of Bristol, HH Wills Physics Laboratory, Tyndall Avenue, Bristol, UK}

\author[0000-0003-0854-3002]{Am\'{e}lie Gressier}
\affiliation{Space Telescope Science Institute, 3700 San Martin Drive, Baltimore, MD 21218, USA}

\author[0009-0002-9017-3679]{Shelby Courreges}
\affiliation{University of Texas at Austin, Department of Astronomy, 2515 Speedway C1400, Austin, TX 78712, USA}

\author[0000-0002-7352-7941]{Kevin B. Stevenson}
\affiliation{Johns Hopkins APL, 11100 Johns Hopkins Rd, Laurel, MD 20723, USA}


\author[0000-0002-5147-9053]{Sukrit Ranjan}
\affiliation{University of Arizona, Lunar and Planetary Laboratory/Department of Planetary Sciences, Tucson, 85721, USA}

\author[0000-0001-8020-7121]{Knicole Col\'{o}n}
\affiliation{NASA Goddard Space Flight Center, Greenbelt, MD 20771, USA}

\author[0000-0003-2528-3409]{Brett M. Morris}
\affiliation{Space Telescope Science Institute, 3700 San Martin Drive, Baltimore, MD 21218, USA}

\author[0000-0003-4816-3469]{Ryan J. MacDonald}
\affiliation{Department of Astronomy, University of Michigan, 1085 S. University Ave., Ann Arbor, MI 48109, USA}
\affiliation{NHFP Sagan Fellow}

\author[0000-0002-2508-9211]{Douglas Long}
\affiliation{Space Telescope Science Institute, 3700 San Martin Drive, Baltimore, MD 21218, USA}

\author[0000-0003-4328-3867]{Hannah R. Wakeford}
\affiliation{University of Bristol, HH Wills Physics Laboratory, Tyndall Avenue, Bristol, UK}


\author[0000-0003-3305-6281]{Jeff A. Valenti}
\affiliation{Space Telescope Science Institute, 3700 San Martin Drive, Baltimore, MD 21218, USA}

\author[0000-0001-8703-7751]{Lili Alderson}
\affiliation{Department of Astronomy, Cornell University, 122 Sciences Drive, Ithaca, NY 14853, USA}

\author[0000-0003-1240-6844]{Natasha E. Batalha}
\affiliation{NASA Ames Research Center, MS 245-3, Moffett Field, CA 94035, USA}

\author[0000-0002-8211-6538]{Ryan C. Challener}
\affiliation{Department of Astronomy and Carl Sagan Institute, Cornell University, 122 Sciences Drive, Ithaca, NY 14853, USA}

\author[0000-0001-5732-8531]{Jingcheng Huang}
\affiliation{Department of Earth, Atmospheric and Planetary Sciences, Massachusetts Institute of Technology, Cambridge, MA 02139, USA}

\author[0000-0003-0525-9647]{Zifan Lin}
\affiliation{Department of Earth, Atmospheric and Planetary Sciences, Massachusetts Institute of Technology, Cambridge, MA 02139, USA}

\author[0000-0002-2457-272X]{Dana R. Louie}
\affiliation{Catholic University of America, Department of Physics, Washington, DC, 20064, USA}
\affiliation{Exoplanets and Stellar Astrophysics Laboratory (Code 667), NASA Goddard Space Flight Center, Greenbelt, MD 20771, USA}
\affiliation{Center for Research and Exploration in Space Science and Technology II, NASA/GSFC, Greenbelt, MD 20771, USA}

\author[0000-0003-0814-7923]{Elijah Mullens}
\affiliation{Department of Astronomy and Carl Sagan Institute, Cornell University, 122 Sciences Drive, Ithaca, NY 14853, USA}

\author[0000-0002-2643-6836]{Daniel Valentine}
\affiliation{University of Bristol, HH Wills Physics Laboratory, Tyndall Avenue, Bristol, UK}

\author{C. Matt Mountain}
\affiliation{Association of Universities for Research in Astronomy, 1331 Pennsylvania Avenue NW Suite 1475, Washington, DC 20004, USA}

\author{Laurent Pueyo}
\affiliation{Space Telescope Science Institute, 3700 San Martin Drive, Baltimore, MD 21218, USA}

\author[0000-0002-3191-8151]{Marshall D. Perrin}
\affiliation{Space Telescope Science Institute, 3700 San Martin Drive, Baltimore, MD 21218, USA}

\author[0000-0003-3858-637X]{Andrea Bellini}
\affiliation{Space Telescope Science Institute, 3700 San Martin Drive, Baltimore, MD 21218, USA}

\author{Jens Kammerer}
\affiliation{European Southern Observatory, Karl-Schwarzschild-Straße 2, 85748 Garching, Germany}

\author[0000-0001-9673-7397]{Mattia Libralato}
\affiliation{INAF-Osservatorio Astronomico di Padova, Via dell'Osservatorio 5, 35122 Padova, Italy}

\author{Isabel Rebollido}
\affiliation{European Space Agency (ESA), European Space Astronomy Centre (ESAC), Camino Bajo del Castillo s/n, 28692 Villanueva de la
Ca\~nada, Madrid, Spain}

\author{Emily Rickman}
\affiliation{European Space Agency (ESA), ESA Office, Space Telescope Science Institute, 3700 San Martin Drive, Baltimore, MD 21218}

\author[0000-0001-8368-0221]{Sangmo Tony Sohn}
\affiliation{Space Telescope Science Institute, 3700 San Martin Drive, Baltimore, MD 21218, USA}
\affiliation{Department of Astronomy \& Space Science, Kyung Hee University, 1732 Deogyeong-daero, Yongin-si, Gyeonggi-do 17104, Republic of Korea}

\author[0000-0001-7827-7825]{Roeland P. van der Marel}
\affiliation{Space Telescope Science Institute, 3700 San Martin Drive, Baltimore, MD 21218, USA}
\affiliation{William H. Miller III Department of Physics and Astronomy, Johns Hopkins University, Baltimore, MD 21218, USA}



\begin{abstract}
TRAPPIST-1~e is one of the very few rocky exoplanets that is both amenable to atmospheric characterization and that resides in the habitable zone of its star --- located at a distance from its star such that it might, with the right atmosphere, sustain liquid water on its surface. Here, we present a set of 4 \textit{JWST}/NIRSpec PRISM transmission spectra of TRAPPIST-1~e obtained from mid to late 2023. Our transmission spectra exhibit similar levels of stellar contamination as observed in prior works for other planets in the TRAPPIST-1 system \citep[e.g.][]{lim:2023, radica:2024}, but over a wider wavelength range, showcasing the challenge of characterizing the TRAPPIST-1 planets even at relatively long wavelengths (3-5 $\mu$m). While we show that current stellar modeling frameworks are unable to explain the stellar contamination features in our spectra, we demonstrate that we can marginalize over those features instead using Gaussian Processes, which enables us to perform novel exoplanet atmospheric inferences with our transmission spectra. In particular, we are able to rule out cloudy, primary H$_2$-dominated ($\gtrsim$ 80$\%$ by volume) atmospheres at better than a 3$\sigma$ level. Constraints on possible secondary atmospheres on TRAPPIST-1~e are presented in a companion paper \citep{Glidden2025}. Our work showcases how \textit{JWST} is breaking ground into the precisions needed to constrain the atmospheric composition of habitable-zone rocky exoplanets. 

\end{abstract}



\section{Introduction} \label{sec:intro}

In over two years of scientific operations, \textit{JWST} \citep{jwst} has revolutionized the field of exoplanet atmospheres. Its unprecedented spectrophotometric precision and wavelength coverage allow it to routinely detect water, carbon dioxide and even products of photochemistry on gas giant exoplanet atmospheres \citep[see, e.g., ][and references therein]{cartermay}, detect rich inventories of carbon-bearing species on sub-Neptunes for the first time \citep[see, e.g.,][]{madhu:2023, benneke:2024, holmberg:2024, beatty:2024}, and very recently even constrain the 
atmospheric composition --- or lack thereof --- of rocky exoplanets \citep[e.g.,][]{greene:2023, zieba:2023, lim:2023, lhs475, kirk:2024, zhang:2024, radica:2024, maymac:2023, ms:2023, alderson:2024, hu:2024, xue:2024, patel:2024, gressier:2024, scarsdale:2024, wachiraphan:2024, alam:2024, august:2024}.

Among the rocky planetary systems amenable for atmospheric characterization with \textit{JWST}, the seven Earth-sized planet system orbiting TRAPPIST-1 arguably stands as one of the most exciting to perform such detailed studies \citep{gillon:2017}. The system provides unique avenues for characterization as the M-dwarf they orbit is small, which maximizes the observability of the signals in both emission and transmission spectroscopy \citep[see, e.g., the discussion in][]{doyon:2024}. Perhaps most interestingly, however, the seven planets span a wide range of instellations (the amount of energy incident on a planetary body from its host star), with the inner planets TRAPPIST-1~b, c and d having $\times$4, $\times$2 and $\times$1.1 times Earth's instellation, TRAPPIST-1~e, f and g spanning the system's Habitable Zone (HZ) of their stars --- the distance at which exoplanets with the right atmospheres might sustain liquid water on their surfaces \citep{kasting:1993, kopparapu:2013} --- and TRAPPIST-1~h having only 10\% Earth's instellation. While it is heavily debated in the literature whether these exoplanets should have retained atmospheres at all given the large expected cumulative XUV irradiation produced by their active M-dwarf host \citep[see, e.g.,][and references therein]{garraffo:2017, wheatley:2017, turbet:2020, kt:2023, looveren:2024}, the system's wide range of instellations provides a unique laboratory to test these predictions --- ultimately offering an opportunity to test whether the Solar System's Cosmic Shoreline, which would predict atmospheres to be more likely for the rocky exoplanets farther away from the star and with larger escape velocities, stands also for planets orbiting stars elsewhere \citep{cosmicshoreline}.

While pioneering observations with \textit{HST}/WFC3 have provided some constraints on possible atmospheric compositions for all the planets in the system \citep[see, e.g.,][]{dewit:2016, dewit:2018, zhang:2018, garcia:2022, gressier:2022}, recent \textit{JWST} exploration through different instrument modes have provided even tighter constraints thanks to the improved precision and wavelength coverage of the observatory. Initial secondary eclipse observations with \textit{JWST}/MIRI 15 $\mu$m photometry \citep{greene:2023, zieba:2023} and $0.6-2.8\ \mu$m transmission spectroscopy with NIRISS/SOSS \citep{lim:2023, radica:2024} have both revealed that sizeable atmospheres for the TRAPPIST-1~b and c planets are unlikely. While these results put little constraints on the kind of atmospheres that the exoplanets farther away from their host star have \citep{kt:2023, gialluca:2024}, the NIRISS/SOSS results highlight how stellar contamination --- the distortion of the observed transmission spectrum due to stellar surface heterogeneities \citep[see, e.g.,][and references therein]{rackham:2018} --- largely shapes the observed transmission spectra at least at wavelengths $<3\ \mu$m, showcasing it as the biggest challenge when it comes to studying the exoplanets in the system. Whether this holds true for wavelengths longer than 3 $\mu$m, where some predictions suggest stellar contamination could be less prominent \citep[see, e.g.,][]{rackham:2018}, has not been studied empirically to date.

Here we present a set of 4 transmission spectra for TRAPPIST-1~e, one of the exoplanets in the habitable zone of the TRAPPIST-1 system ($R_p = 0.92R_\oplus$, $M_p = 0.69M_\oplus$, $P = 6.10$ days, $T_{\textrm{eq}} = 250$ K) obtained with \textit{JWST}/NIRSpec PRISM on the $0.6-5\ \mu$m range. This work is part of a series of studies being pursued by the \textit{JWST} Telescope Scientist Team (\textit{JWST}-TST)\footnote{\label{fn-tst-team} \url{https://www.stsci.edu/~marel/jwsttelsciteam.html}}, which uses Guaranteed Time Observations (GTO) time awarded by NASA in 2003 (PI M. Mountain) for studies in three different subject areas: (a) Transiting Exoplanet Spectroscopy (lead: N. Lewis); (b) Exoplanet and Debris Disk High-Contrast Imaging (lead: M. Perrin); and (c) Local Group Proper Motion Science (lead: R. van der Marel). A common theme of these investigations is the desire to pursue and demonstrate science for the astronomical community at the limits of what is made possible by the exquisite optics and stability of \textit{JWST}. The present paper is part of our work on transiting exoplanet spectroscopy, which focuses on Deep Reconnaissance of Exoplanet Atmospheres using Multi-instrument Spectroscopy \citep[DREAMS; see, e.g.,][]{dreams1, dreams2, dreams3, Louie_2025} of three transiting exoplanets representative of key classes: Hot Jupiters (WASP-17b, GTO~1353), Warm Neptunes (HAT-P-26b, GTO~1312), and Temperate Terrestrials (TRAPPIST-1e, GTO~1331).

Our work is divided in two papers. The current paper presents the \textit{JWST} NIRSPec/PRISM data reduction and analysis, as well as the necessary methodologies to interpret our transmission spectra, which as we show throughout this work we believe is dominated by stellar contamination. A companion paper \citep{Glidden2025} provides a deep-dive on the implications of our atmospheric constraints for possible secondary atmospheres on TRAPPIST-1~e. This current paper is divided as follows. In Section \ref{sec:observations} we introduce our observations and data reduction framework. In Section \ref{sec:results} we present our results from modelling both the white-light and spectrophotometric light curves, as well as the retrieval framework used in our work to extract inferences from possible planetary atmospheres on TRAPPIST-1~e, which introduces a novel Gaussian Process framework to deal with stellar contamination. Section \ref{sec:discussion} offers a discussion of our results, contextualizing them with previous constraints and observations. Our final conclusions are presented in Section \ref{sec:conclusions}.

\section{Observations and Data Reduction} \label{sec:observations}

\subsection{Observations}

\begin{figure*}
\includegraphics[width=2.1\columnwidth]{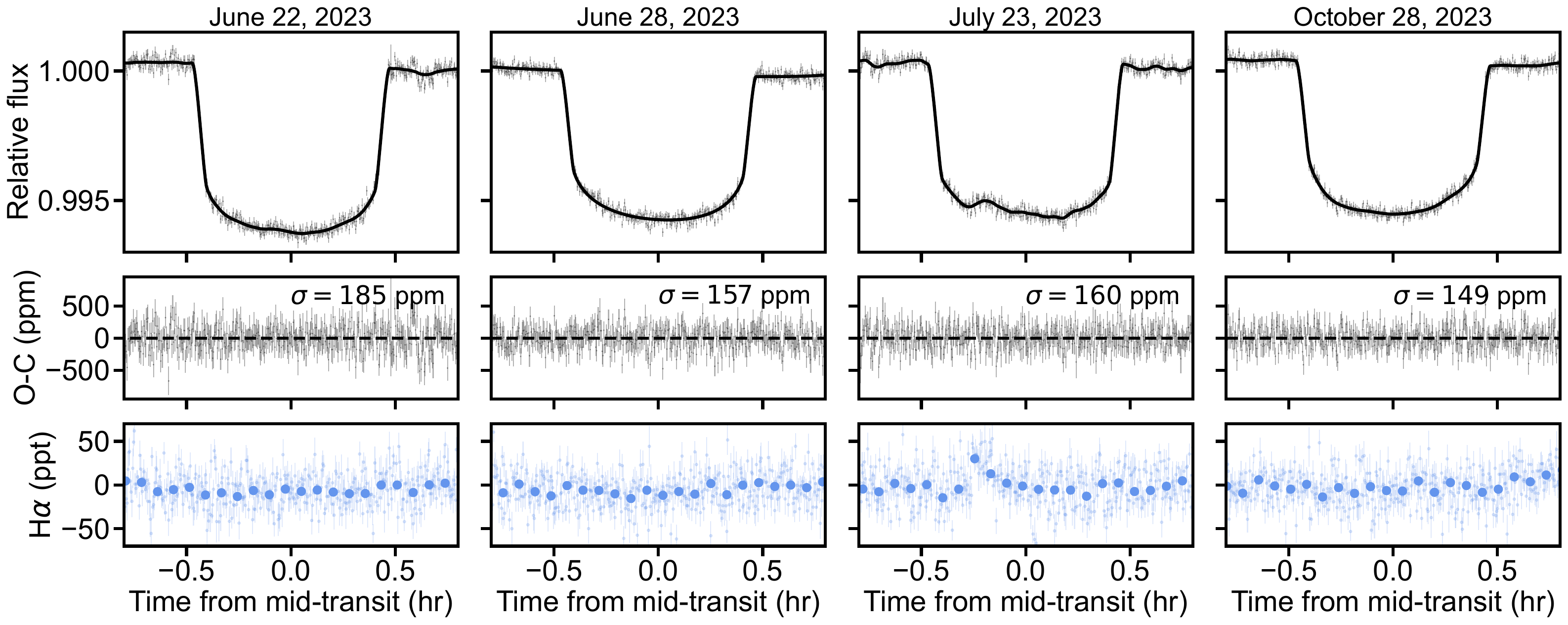}
\caption{\textbf{White-light TRAPPIST-1~e \textit{JWST}/NIRSpec PRISM transit light curves}. (Top) Datapoints of the transit event (grey; binned at a cadence of 14-seconds) along with the best-fit transit plus systematics model (black; which includes a visit-long slope and a GP; see text for details). The date at which each observation was obtained is indicated at the top of each panel. (Middle) Residuals of the data minus the best-fit light curve model in parts-per-million (ppm); the root-mean-square of the data at this 14-second cadence is indicated for each observation. (Bottom) Light curves at the H$\alpha$ line ($0.656 \pm 0.02\ \mu$m, in parts per thousand; ppt) at the same cadence as the white-light curves (pale blue points), and binned at 5-minutes (solid blue points); note how a flare is revealed at about $-0.25$ hr in this light curve as the most likely explanation for the bump in the July 23, 2023 transit event.
\label{fig:white-light}}
\end{figure*}

Observations targeting a primary transit of TRAPPIST-1~e with \textit{JWST}/NIRSpec PRISM were obtained by GTO~1331 \citep{t1eproposal} on June 22, June 28, July 23 and October 28, 2023. Target Acquisition was performed on each visit on TRAPPIST-1 itself. These 4 transit observations consisted of 4-hour, 5-groups per integration exposures using the \texttt{SUB512} subarray, which gave a cadence of 1.38 seconds per integration and allowed plenty of (a) time-baseline to observe the $\sim 1$-hour transit events and (b) non-illuminated detector space to subtract possible background signals and other detector systematics. While this setup allowed us to obtain spectrophotometry in a wide wavelength range (0.6-5 $\mu m$), it did make some pixels in the $1.1-1.7\ \mu$m region reach fluence levels above 90\% the saturation level of the detector in the last group, which, as has been discussed in the literature, requires special care \citep{w39prism, cartermay}. 

\subsection{Data Reduction} \label{sec:datareduction}

The data were reduced with a variety of pipelines to validate the robustness of the signals observed in our transmission spectra --- the details of this comparison and of each data reduction pipeline are detailed in Appendix \ref{sec:red}. In what follows, we make use of the results obtained by using the \texttt{transitspectroscopy} pipeline \citep{transitspectroscopy}, which are detailed in Appendix \ref{ne-reduction}. In short, this mainly makes use of the \texttt{JWST Calibration Pipeline}'s Stage 1 to perform detector calibration of \texttt{*uncal.fits} files all the way to the rates per integration \citep{jwstcal}; the only different step is a custom jump detection algorithm. 1/f noise reduction is performed at the rates per integration level using a similar methodology as the one introduced in \cite{radica:2023} for NIRISS/SOSS. The spectral traces of each integration are obtained via cross-correlation at each column with a Gaussian, whose peaks are fitted by a B-spline to smooth the trace at each integration. Spectral extraction is performed via simple extraction. 

Light curves are then fitted with the \texttt{juliet} library \citep{juliet}. For the band-integrated, white-light light curves, we let the limb-darkening coefficients be free parameters using a quadratic limb-darkening law{, where we use the \cite{kipping2013} parametrization for limb-darkening. While as noted in \cite{coulombe:2024} in general this parametrization might introduce biases for JWST-quality light curves, we believe this is not important in our case as the band-integrated flux for NIRSpec/PRISM is dominated by flux $< 2 \ \mu$m where this same work shows this effect is minimal}. However, for ease of comparison across data reduction pipelines, we decided to fix the limb-darkening coefficients in the wavelength-dependent light curve fits to the ones predicted by PHOENIX models using the \texttt{limb-darkening} library \citep{ld}. A Gaussian Process \citep[GP;][]{gps:2023} is used to model systematic trends in the observed light curves{. The band-integrated and wavelength dependent light curves were fitted with 3 possible kernels; an exponential, a Mat\`ern 3/2 and an exponential-Mat\`ern 3/2 kernel (i.e., a multiplication of both). We found identical results with the three; here we show the results for the} Mat\`ern 3/2 kernel to parameterize the GP using the \texttt{celerite} library \citep{celerite}. The use of GPs produces, overall, slightly larger error bars on the estimated transmission spectrum for the four visits, which is one of the reasons why we decide to use this pipeline to perform inferences on, as it gives rise to overall more conservative error bars on the final transmission spectrum (although, as described in Appendix \ref{sec:red}, with variations seen as a function of wavelength consistent within reductions for the 4 visits).

\section{Results} \label{sec:results}

\subsection{White-light light curve analysis} \label{sec:white-light}

Figure \ref{fig:white-light} showcases a close-up to the observed transit events of TRAPPIST-1~e on the white-light light curves for each of our observation dates using the data reduction procedure described above as grey points, in order to showcase the overall data quality of our observations. The priors and posteriors used for each individual light curve fit (i.e., each visit) are presented in Table \ref{tab:wl-params}, and the best-fit light curves are presented in black along with the corresponding residuals in Figure \ref{fig:white-light}.

\begin{deluxetable*}{lcccccc}[t]
\tablecaption{Prior and posterior parameters of the white-light (i.e., band-integrated) lightcurve fits performed on the NIRSpec/PRISM data of TRAPPIST-1~e. For the priors, $N(\mu,\sigma^2)$ stands for a normal distribution with mean $\mu$ 
and variance $\sigma^2$; $U(a,b)$ stands for a uniform distribution between $a$ and $b$, respectively and $\log U(a,b)$ stands for a log-uniform prior on the same range. Priors for times of transit $T_{pred}$, period $P$, impact parameter $b$ and semi-major axis to stellar radius ratio $a/R_*$ come from \cite{agol-t1}. Priors are for the individual light curve fits of each visit. Combined posterior (last column) corresponds to the average of the 4 independent fits.\label{tab:wl-params}}
\tablecolumns{4}
\tablewidth{0pt}
\tablehead{
\colhead{Parameter} &
\colhead{Prior} &
\colhead{June 22} &
\colhead{June 28} &
\colhead{July 23} &
\colhead{October 28} & 
\colhead{Combined}
}
\startdata
\multicolumn{5}{l}{\textit{Physical \& orbital parameters}} \\
\ \ \ \ $P-P_0$ (sec)$^{\psi}$ & $N(0,3^2)$ & $-0.12^{+3.02}_{-2.85}$ & $-0.20^{+3.06}_{-3.01}$ & $0.00^{+3.03}_{-3.01}$ & $-0.06^{+2.99}_{-2.99}$ & $-0.09^{+1.47}_{-1.47}$ \\
\ \ \ \ $T_{0}-T_{pred}$ (sec)$^{\psi}$ & $N(0,8640^2)$ & $-82.17^{+2.55}_{-2.76}$ & $-75.24^{+1.91}_{-1.79}$ & $-47.63^{+3.69}_{-3.48}$ & $95.92^{+2.48}_{-2.60}$ & ---\\
\ \ \ \ $\left(R_{p}/R_\star\right)^2$ (ppm)  &  $U(0.0,0.2)$ on $R_p/R_*$ & $5626^{+90}_{-90}$ & $4850^{+73}_{-66}$ & $5150^{+99}_{-99}$ & $4907^{+78}_{-88}$  & $5133^{+45}_{-46}$  \\
\ \ \ \ $b=(a/R_\star)\cos(i)$  &  $N(0.191,0.041^2)$ & $0.156^{+0.030}_{-0.031}$ & $0.168^{+0.028}_{-0.031}$ & $0.183^{+0.028}_{-0.031}$ & $0.186^{+0.027}_{-0.030}$ &  $0.172^{+0.014}_{-0.015}$  \\
\ \ \ \ $a/R_\star$  &  $N(52.86,0.39^2)$ & $52.58^{+0.22}_{-0.25}$ & $52.65^{+0.24}_{-0.27}$ & $52.82^{+0.27}_{-0.28}$ & $52.84^{+0.27}_{-0.28}$ &  $52.72^{+0.13}_{-0.13}$ \\
\ \ \ \ $e$  &  fixed &  0 & 0 & 0 & 0 & 0 \\
\ \ \ \ $\omega$ (deg) &  fixed & 90 & 90 & 90 & 90 & 90 \\
\multicolumn{2}{l}{\textit{Limb-darkening coefficients}} \\
\ \ \ \ $u_{1}^\dagger$ &  $U(0,1)$ on $q_1^\dagger$ & $0.25^{+0.11}_{-0.11}$ & $0.283^{+0.056}_{-0.057}$ & $0.17^{+0.14}_{-0.11}$ & $0.33^{+0.11}_{-0.11}$ &  $0.261^{+0.053}_{-0.049}$ \\
\ \ \ \ $u_{2}^\dagger$ &  $U(0,1)$ on $q_2^\dagger$ & $0.24^{+0.15}_{-0.17}$ & $0.285^{+0.092}_{-0.084}$ & $0.27^{+0.15}_{-0.19}$ & $0.32^{+0.15}_{-0.15}$ & $0.275^{+0.070}_{-0.074}$ \\
\multicolumn{2}{l}{\textit{Instrument systematics}} \\
\ \ \ \ $M$ (ppm)$^{\alpha}$ &  $N(0,10^5)$ & $-183^{+79}_{-78}$ & $-45^{+308}_{-620}$ & $-218^{+68}_{-69}$ & $-331^{+78}_{-68}$ & $-197^{+107}_{-166}$ \\
\ \ \ \ $\sigma_{GP}$ (ppm)$^{\beta}$ &  $\log U(10,10^9)$ & $159^{+69}_{-35}$ & $376^{+1660}_{-211}$ & $186^{+40}_{-28}$ & $134^{+85}_{-35}$ & $239^{+470}_{-72}$ \\
\ \ \ \ $\rho_{GP}$ (hours)$^{\beta}$ &  $\log U(0.024,2400)$ & $0.16^{+0.19}_{-0.09}$ & $1.35^{+4.04}_{-0.82}$ & $0.075^{+0.028}_{-0.018}$ & $0.18^{+0.13}_{-0.07}$ & $0.48^{+1.05}_{-0.23}$ \\
\ \ \ \ $\sigma_{w}$ (ppm)$^{\beta}$ &  $\log U(10,10^3)$ & $139.00^{+8.31}_{-8.36}$ & $113.91^{+6.86}_{-6.66}$ & $115.89^{+8.15}_{-7.66}$ & $103.90^{+6.56}_{-6.56}$ & $118.24^{+3.69}_{-3.77}$ \\
\enddata
\tablenotetext{\psi}{Here, $P_0 = 6.101013$ days is the best-fit period in \cite{agol-t1}. $T_{pred}$ are the predicted transit times for our visits in \cite{agol-t1}, which are 2460118.460787, 2460124.55984, 2460148.956812 and 2460246.538176 for June 22, June 28, July 23 and October 28, 2023 respectively. All times in BJD TDB.}
\tablenotetext{\dagger}{Quadratic limb-darkening law; priors were set on the $q_1$ and $q_2$ parameters using the transformations in \cite{kipping2013} to obtain $u_1$ and $u_2$.}
\tablenotetext{\alpha}{The transit model is multiplied by $1/(1 + M)$ to account for possible normalization offsets made to the light curves when obtaining relative fluxes. For details see Section 2.1 in \cite{juliet}.}
\tablenotetext{\beta}{Here, $\sigma_{GP}$ and $\rho_{GP}$ represent the amplitude and time-scale of a Mat\`ern 3/2 GP; $\sigma_w$ represents jitter added in quadrature to error bars.}
\end{deluxetable*}

Overall, the white-light curves showcase very high precision, albeit at higher noise levels than what would be expected by the estimated error bars by the \textit{JWST} pipeline \citep[as is true in most \textit{JWST} white-light transit light curves, very likely stemming from residual 1/f noise, see, e.g.,][]{cartermay, Espinoza:2023}. However, it is evident different epochs present different levels of systematic trends. The July 23, 2023 visit, for instance, presents an evident bump just before mid-transit which, following \cite{Howard_2023} and \cite{radica:2023}, we identify as a small flare, judging from the shape of the light curve at H$\alpha$ wavelengths (Figure \ref{fig:white-light}, bottom panel); similar small oscillations are observed in the June 22, 2023 light curve, with the rest of the light curves not showcasing as strong systematic effects. Indeed, this is reflected in the best-fit amplitudes of the GP's ($\sigma_{GP}$ parameter) for each visit presented in Table \ref{tab:wl-params} --- the amplitude is tightly constrained to be about a few hundred ppm for both of those visits, whereas the parameter is loosely constrained for the June 28, 2023 visit (where a lower level of systematic effects are observed) and appears to be slightly smaller for the October 28, 2023 visit. Interestingly, the timescales of the process all appear to be relatively similar, on the order of 5-10 minutes (except, again, for the June 28, 2023 visit where the parameter is also loosely constrained). Investigations on time-series of parameters of the 2D spectra, such as trace motion and profile widths reveal no obvious correlation with the variability observed in our light curves. Similar studies performed on the \textit{JWST} guide star data using the \texttt{spelunker} library \citep{spelunker} provided similar null correlation results. This suggests that the systematic trends observed in our light curves might actually be due to time-varying phenomena produced by TRAPPIST-1 (the star) itself, such as flares or some type of stellar oscillation. While studying the detailed origin of these is outside the scope of this work, we do account for them in our wavelength-dependent light curve fitting via GPs. We did find that other methodologies give rise however to similar results (see Appendix \ref{sec:red} for details on this comparison). 

Another interesting element of the individual visit results is the larger white-light transit depth on the first visit of the program on June 22, 2023. This difference is particularly curious given the fact that the depth difference is the largest when compared against the June 28, 2023 visit --- which happened only 6 days after the June 22 visit \citep[or about 2 stellar rotation periods, considering a rotation period of 3.3 days as measured in][]{morris:2018}. The transit depth difference between those two visits is $774^{+116}_{-120}$ ppm, which is a difference that is significant at more than 5$\sigma$ --- and one that can be visually inspected in Figure \ref{fig:white-light} as well. As it will be shown in the next section, this variability is observed at the same absolute level in \textit{all} of our data reduction methodologies, which strongly suggests that this is real variability in the transit depths --- very likely stemming from stellar contamination \citep{rackham:2018}.

Despite the above effects, our white-light curve results in Table \ref{tab:wl-params} already show a significant improvement in the orbital parameters of TRAPPIST-1~e by a factor of $\sim 3$ in both the impact parameter and the scaled semi-major axis, and by a factor of $\sim 50$ on the predicted time-of-transits from \cite{agol-t1}. As has been shown in \cite{agol:2024}, the latter are of particular importance for improving the overall ephemerides for the planetary system around TRAPPIST-1. This showcases once again that \textit{JWST} white-light curves, although typically byproducts of atmospheric characterization, provide excellent datasets to refine planetary system properties at levels that were unattainable by previous instrumentation \citep[see, e.g.,][and references therein]{mahajan2024, cartermay}.

\subsection{The transmission spectra} \label{sec:transpec}

\begin{figure*}
\includegraphics[width=2.1\columnwidth]{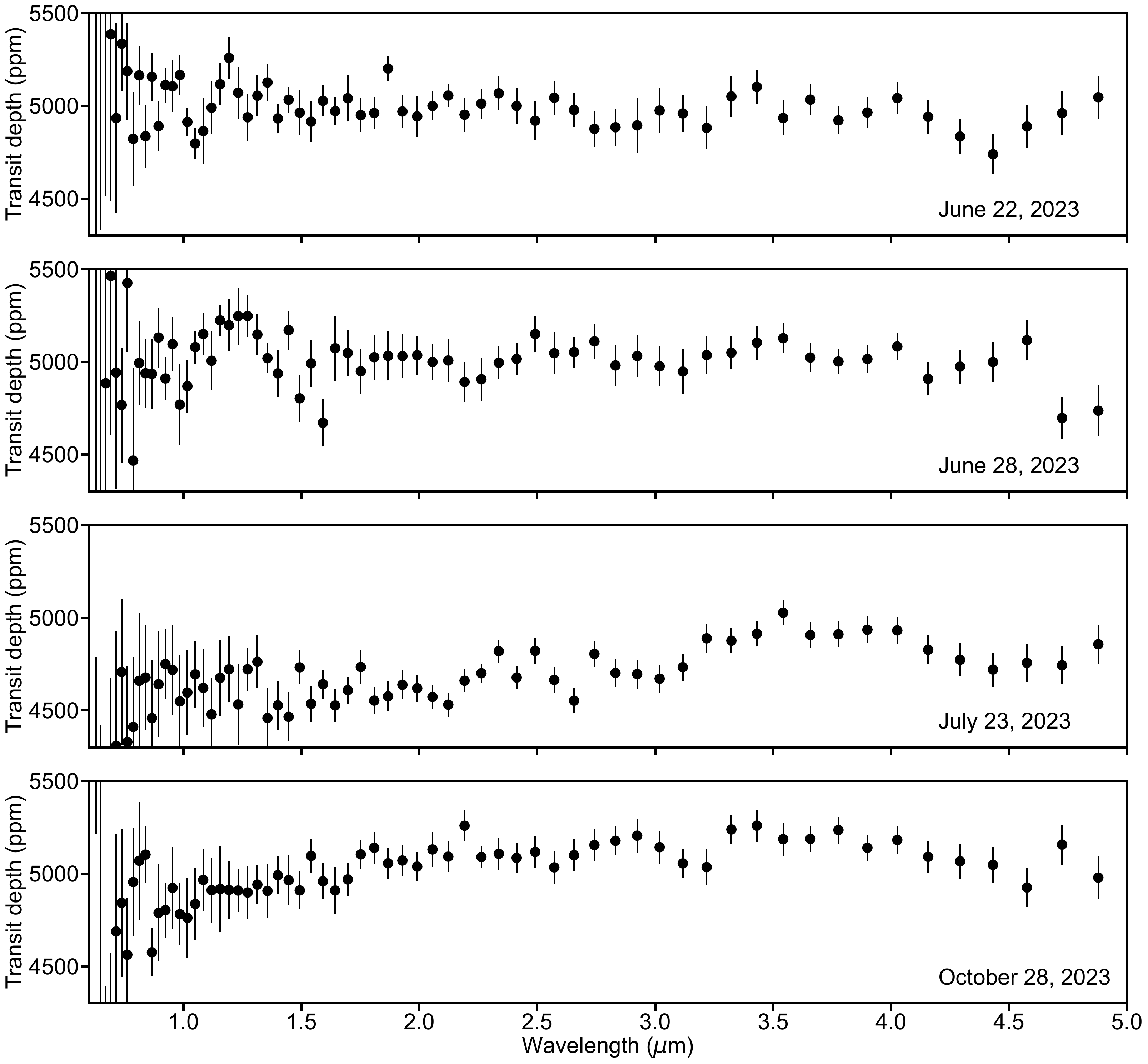}
\caption{\textbf{TRAPPIST-1~e NIRSpec/PRISM transmission spectra on different epochs.} The transmission spectra are ordered in chronological order from top to bottom, with the dates at which they were obtained indicated in the lower right of each panel. The y-axes for all plots cover the same ranges, which highlight how much the transmission spectrum varies across epochs --- in particular for the transmission spectra obtained in July and October 2023. 
\label{fig:tspectra}}
\end{figure*}

The transmission spectra obtained for our different observing dates are presented in Figure \ref{fig:tspectra}. These were obtained by following the same procedures described above for the white-light light curve analysis, but fixing the orbital parameters (i.e., the time-of-transit center, impact parameter, scaled semi major axis, eccentricity and argument of periastron) as well as the limb-darkening coefficients {using a quadratic law}, as described in Section \ref{sec:datareduction}. As can be observed, the data showcases an overall flatter transmission spectrum from 0.6-5 $\mu$m on the June 2023 visits, and strong increases in transit depth towards longer wavelengths for the visits of 
July and October 2023. As we show in Appendix \ref{sec:red}, these variations in the transmission spectra for different epochs are observed in \textit{all} of our reductions, which use different pipelines to reduce the raw data and different methodologies to fit the transit light curves themselves, which showcases the robustness of the observed features in the transmission spectra. 

Given TRAPPIST-1 is well-known as a magnetically active star, shown to frequently flare and host both hot and cold spots on the stellar surface through evidence in photometric monitoring campaigns and transmission spectra \citep[see e.g.][]{morris:2018,Wakeford_2019,Ducrot_2020,lim:2023, Howard_2023, radica:2024}, we interpret those epoch-to-epoch variations as a function of wavelength in our observed transmission spectra as mostly being the product of stellar heterogeneities contaminating our transmission spectra, which are evolving in time. 

Initial attempts at using existing retrieval analysis tools that include stellar contamination modeling to explain our observed transmission spectra were unsuccessful at reproducing their complex observed variation as a function of wavelength. As we show in Appendix \ref{sec:stellarcontamination}, while publicly available stellar models were adequate to model the visits in June 2023, from which we infer the possible existence of heterogeneities colder than the stellar photosphere (i.e., ``cold" spots), they are unsuccessful at modeling the visits on July 23, 2023 and October 28, 2023, both of which showcase strong evidence for hot stellar heterogeneities (i.e., ``hot" spots).  Those experiments suggest that, although our data presents strong evidence for stellar contamination in the transmission spectrum of TRAPPIST-1~e, performing inferences on it is not straightforward with current stellar models and/or retrieval frameworks. It is very likely this is due to the limitations of using 1D radiative/convective equilibrium stellar atmosphere models with different temperatures as a proxy for the spectral features of cold and hot ``spots", which are known to arise in magnetically active regions and thus give rise to very complex emergent fluxes \citep[see e.g.][]{Witzke_2022, norris:2023, Smitha_2025}. We thus decided to develop a new methodology to perform joint stellar contamination and atmospheric retrievals on our observed TRAPPIST-1~e transmission spectra using GPs to incorporate our limited knowledge on the underlying data generating process.

\subsubsection{Atmospheric inferences using GPs}

The complex structure of our observed transmission spectra points to opacity sources and/or physical mechanisms defining the emergent spectra of cold and hot ``spots" ---or even the photospheres of M-dwarfs--- that are not included in current publicly available stellar models \citep[such as, e.g., magnetic field impacts on the spectra of ``hot" spots;][]{norris:2023} or even as of yet unidentified systematic effects impacting all of our data reduction pipelines. This makes inferences on the possible atmospheric composition of TRAPPIST-1~e from our observed transmission spectra not straightforward to perform with either forward models, model grids or atmospheric retrievals, as all of those that include models for stellar contamination parametrize its impact using models similar to the ones discussed above \citep[see, e.g.,][and references therein]{retrievals:2023}.

At the low resolutions we are dealing with in this work, both stellar and exoplanetary spectra are expected to be relatively smooth functions with well defined length-scales. Motivated by this assumption, along with the versatility of GPs to model and account for unknown systematic trends in transit light curves with such properties, here we decide to model the unknown signals that distort our transmission spectra using these processes as well. The framework for incorporating GPs in atmospheric retrievals has already been introduced in the literature for transmission spectra in the case of additive signals that distort the spectra due to unknown systematic effects \citep[see, e.g.,][]{guilluy:2024,mccreery:2025,rotman:2025}. The difference in incorporating it in our work, however, is that here we are interested in modeling a signal that is \textit{multiplicatively} distorting our transmission spectra, as stellar contamination acts at first order multiplicatively on a transmission spectrum \citep[see, e.g.,][and references therein]{rackham:2018}. Multiplicative GPs, however, are not straightforward to derive. To circumvent this problem, instead of modelling the observed transit depth of a visit $v$, $\delta_{v,i}(\lambda_i)$ at the $i$-th wavelength bin $\lambda_i$ we model the \textit{logarithm} of this transit depth as

\begin{widetext}
\begin{equation}
    \label{eq:log}
    \log \delta_{v,i}(\lambda_i)  \sim \log  \epsilon_{c,v}(\lambda_i) + \log \left[C_v + S(\lambda_i)\right] + \mathcal{GP}_{v} + \epsilon_{w,v}.
\end{equation}
\end{widetext}

This converts our multiplicative problem into an additive one. Here, $\epsilon_{c,v}(\lambda_i)$ is the (visit-dependent, deterministic) stellar contamination signal that distorts the exoplanet atmospheric signal $S(\lambda_i)$, $C_v$ is a (visit-dependent) constant transit depth offset injected by the unknown zero-point of the stellar photosphere, $\mathcal{GP}_v$ represents a GP {for visit $v$}; and $\epsilon_{w,v}$ is a white-noise component that incorporates both the observed transit depth errors at each $\lambda_i$, but also allows for a jitter term $\sigma_w$ added in quadrature to those error bars to account for possible underestimated error bars on the observed transit depths. It is important to note that this methodology of modelling observables in logarithm to include multiplicative signals --- either GPs or linear models --- is not new, and is already employed in the photometric time-series literature \citep[see, e.g.,][and references therein]{Espinoza:2019, Weaver:2020, Allen:2022}.

We provide the details of our GP atmospheric retrieval methodology ---including the priors used in our retrievals--- in Appendix \ref{sec:gp-retrieval-framework}. In short, our retrieval framework uses dynamic nested sampling via the \texttt{dynesty} library \citep{dynesty} to explore the parameter space, using the \texttt{POSEIDON} library \citep{poseidon1, poseidon2} to perform the radiative transfer and compute forward models $S(\lambda_i)$. Following \cite{Lin:2021} and \cite{LY:2023}, we consider H$_2$, CO$_2$, CH$_4$, H$_2$O, N$_2$, O$_2$, O$_3$, N$_2$O and CO as the spectrally active species in this modeling framework. We follow \cite{bennekeseager:2012} and use a centered-log-ratio transformation to model the mixing ratios in order to consider any combination of the spectrally active molecules in our retrievals to be the background gas. We include the impact of clouds via a cloud-top pressure parameter, which we here interpret as an effective surface pressure for the atmospheres under study --- we also allow for the reference pressure to be a free parameter in our retrievals. The deterministic part of the stellar contamination signal,  $\epsilon_{c,v}(\lambda_i)$, is built following the formalism in \cite{rackham:2018}, for which we use BT-SETTL models \citep{bt-settl} to incorporate both hot and cold stellar heterogeneities. For our GP (which we interpret as the stochastic part of our stellar contamination model), we use a Mat\`ern 3/2 kernel via the \texttt{george} library \citep{george}. { We also experimented using a squared-exponential kernel, finding the very same results we showcase below with the Mat\`ern 3/2 kernel.}

\subsection{GP retrieval results}

\begin{figure*}
\includegraphics[width=\textwidth]{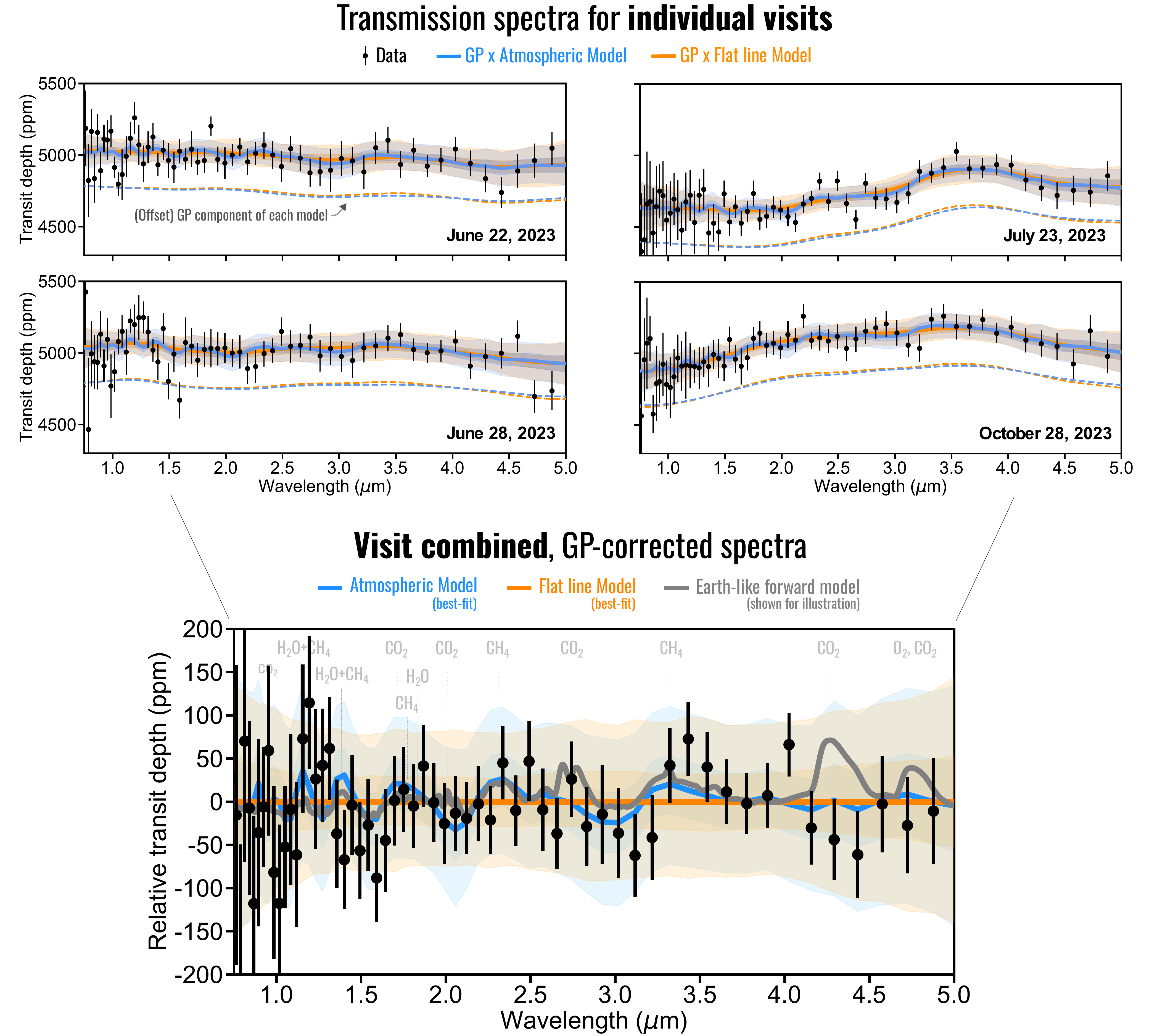}
\caption{\textbf{The transmission spectra of TRAPPIST-1~e interpreted with Gaussian Processes and atmospheric/atmosphereless models.} (\textit{Top}) Transmission spectra on our four visits (black points with error bars) modeled with a GP times either an atmospheric model (blue) or a flat line spectrum (i.e., with no atmosphere or with a high-altitude cloud deck; orange); a GP (offset, in dashed lines) acts multiplicatively to distort those signals. Bands represent the 1 and 3-sigma credibility bands. (\textit{Bottom}) Visit combined transmission spectrum by {(weighted)} averaging the four visits after correcting for the modeled GP component ({using the flat-line model-derived GP;} black points with error bars). The atmospheric model and the flat line model are indistinguishable according to the Bayesian evidence --- more data is needed to distinguish between those. Bands represent the 1 and 3-sigma credibility bands. Note how, within the error bars, an Earth-like model (grey; with the location of main active spectroscopic features) is still consistent with our data. {Also note the blue and orange models are shared but fitted to each individual visit.}
\label{fig:gpretrievals}}
\end{figure*}

We explored performing retrievals of different complexity when it came to defining the stellar contamination model in our framework. We tried performing two-component (i.e., a spot and a photosphere) and three-component (i.e., a ``hot" spot, a ``cold" spot and a photosphere) stellar contamination models, as well as including or not the GP components. We also tried models in which instead of having a different stellar contamination model, a global stellar contamination model --- common to all visits --- was defined. From all those combinations, the models with the largest log-evidences (which are models that have $\Delta \log Z > 5.8$ when compared to all the other model combinations) were two models that set the ``deterministic" stellar contamination model $\epsilon_{c,v}$ to $1$ (i.e., no deterministic stellar contamination component), and which absorb the large, hundreds of ppm variations likely coming from stellar contamination in our observed transmission spectra into the GP. The first was a model with a GP term per visit and an exoplanet atmosphere on TRAPPIST-1~e, while the second was a model with a GP term per visit with a featureless spectrum per visit (i.e., with $S(\lambda_i) = 0$ in our notation above). Both models and the corresponding combined, corrected transmission spectrum by the GP components are presented in Figure \ref{fig:gpretrievals}. 

One of the striking features of the GP corrected spectra presented in Figure \ref{fig:gpretrievals} is the precision that we achieve in our four visit combined {(via a weighted mean)} transmission spectrum (bottom panel) --- we are able to unveil a spectrum with error bars on the order of 50 ppm at $R=30$ in the 0.6-5 $\mu$m range, which significantly expands both in precision and wavelength prior HST/WFC3 constraints on the transmission spectrum of TRAPPIST-1~e \citep[from, e.g.,][]{dewit:2018, zhang:2018}. The difference in log-evidence between models that include exoplanetary atmospheric features (blue model in Figure \ref{fig:gpretrievals}) and those that don't (orange model) is $\Delta \log Z = 5.3$ in favor of the no-atmosphere model using all the molecules (27 total free parameters). However, reducing the network to only H$_2$, CO$_2$, CH$_4$, H$_2$O and CO (the molecules we expect to show the highest impact on our NIRSpec/PRISM transmission spectra) raises the bayesian evidence of the atmospheric model to a level that makes it indistinguishable from the featureless model ($\Delta \log Z = 1.6$ in favor of the featureless model). Based on those analyses, we find that given our current data we are unable to distinguish between models containing exoplanet atmospheric features and models that are featureless.

\subsection{Constraints on possible atmospheric compositions of TRAPPIST-1~e}

\begin{figure}
\includegraphics[width=0.46\textwidth]{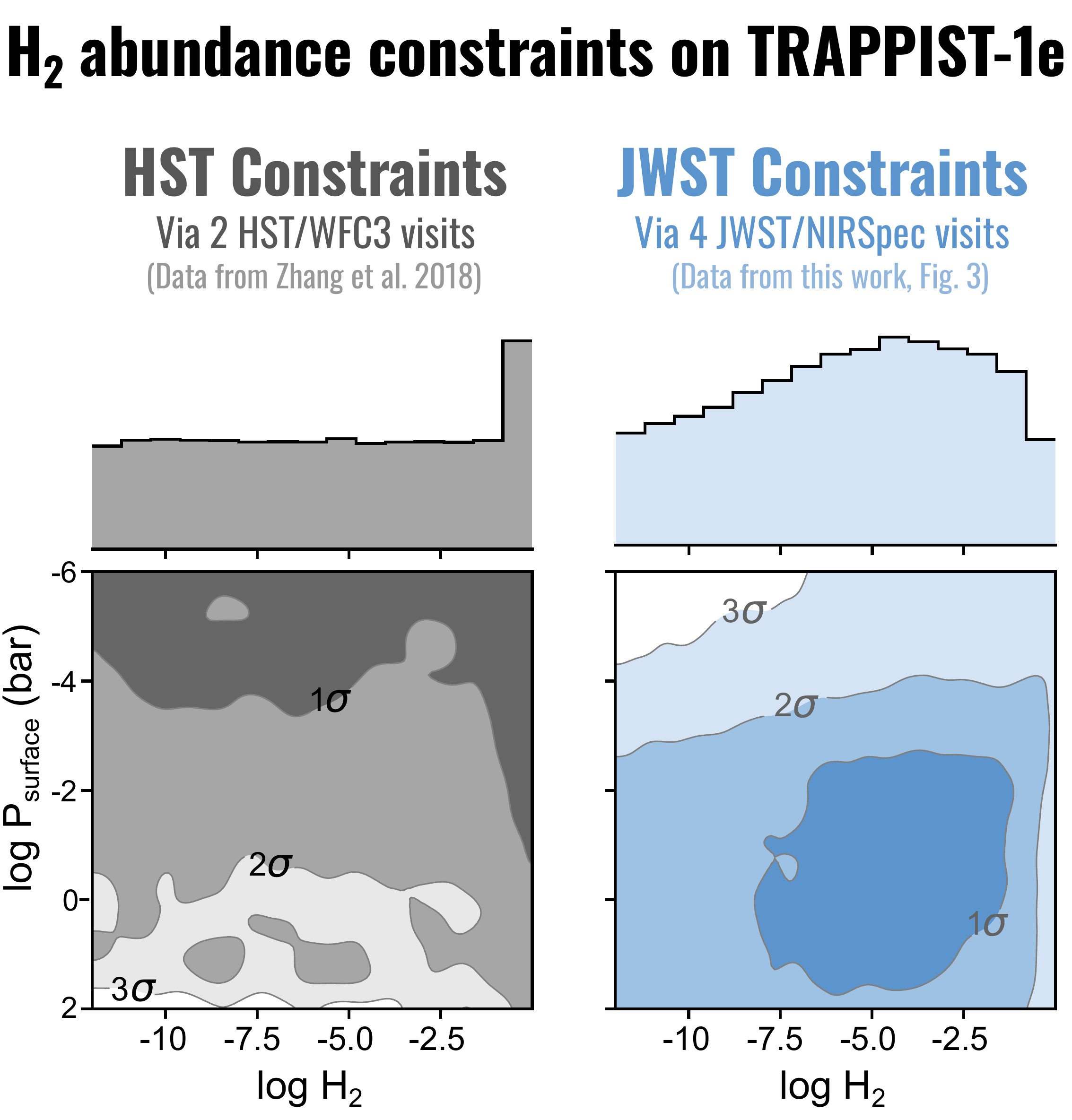}
\caption{\textbf{H$_2$ abundance constraints for TRAPPIST-1~e from HST and JWST as a function of surface pressure.} Posterior distribution showcasing the improvement on constrains on possible H$_2$-dominated atmospheres on TRAPPIST-1~e between \textit{HST} (left in grey, obtained applying our GP-retrieval methodology to the HST/WFC3 data in \citet{zhang:2018}) and \textit{JWST} (right in blue, obtained by applying it to the 4 NIRSpec/PRISM transits presented in this work). The distribution for \textit{HST} mainly follows the centered log-ratio prior allowing the H$_2$-dominated solution at virtually all pressures $\gtrsim$ 1 bar; the \textit{JWST} one disfavors the H$_2$-dominated solution.
\label{fig:h2}}
\end{figure}

Our precise NIRSpec/PRISM spectra, together with our GP retrieval methodology, allow us to put novel constraints on the possible atmospheric compositions of TRAPPIST-1~e. To illustrate the power of these constraints ---and the improvement over previous measurements--- we applied the same GP retrieval methodology introduced above to the previous state-of-the art in transmission spectroscopy for TRAPPIST-1~e: the 2 \textit{HST}/WFC3 visits introduced in \cite{dewit:2018} for this exoplanet, re-analyzed and studied in detail in the work of \cite{zhang:2018}. We provide details of our retrieval framework applied to the \textit{HST}/WFC3 data presented in \cite{zhang:2018} in Appendix \ref{sec:gp-retrieval-framework}. For illustration and comparison, we present constraints on possible primary, H$_2$-dominated atmospheres with our methodology both using these 2 visits and the 4 \textit{JWST}/NIRSpec visits introduced in this work in Figure \ref{fig:h2}.

Studying the \textit{HST}/WFC3 marginal posterior distribution on the H$_2$ abundance constraints for TRAPPIST-1~e (i.e., the grey histogram in Figure \ref{fig:h2}), we are in fact unable to rule out H$_2$-dominated atmospheres at the 3$\sigma$ level. However, we are able to rule out abundances larger than about $80\%$ by volume with our 4 \textit{JWST}/NIRSpec data (blue posterior distributions) even in cloudy/low-pressure scenarios at more than a $3\sigma$ level. Only 1\% of the posterior samples, in fact, allow for H$_2$ abundances larger than $50\%$ by volume when using our \textit{JWST}/NIRSpec data --- making the H$_2$-dominated scenario (and thus primary atmospheres scenarios for TRAPPIST-1~e), very unlikely given our \textit{JWST}/NIRSpec data even in the presence of clouds. A detailed presentation and study of our posterior constraints on possible \textit{secondary} atmospheres for TRAPPIST-1~e as inferred from our GP retrieval framework are presented in a companion paper \citep{Glidden2025}. 

\section{Discussion} \label{sec:discussion}

The \textit{JWST}/NIRSpec PRISM transmission spectrum presented in this work for the habitable-zone exoplanet TRAPPIST-1~e is one of the most precise measurements and constraints on the atmospheric composition of a rocky, habitable-zone exoplanet to date. Among the key lessons from our observations is that stellar contamination --- which distorts our observed transmission spectrum due to unocculted hot and cold ``spots" in the star --- is one of the biggest challenges when it comes to inferring atmospheric properties of the exoplanets in the TRAPPIST-1 system. While this has been shown to be the case as well with NIRISS/SOSS observations at wavelengths $< 3\ \mu$m \citep[see, e.g.,][]{lim:2023, radica:2024}, our work reveals that this might be a problem even for longer wavelengths, where an important number of strong possible absorbers (such as, e.g., CH$_4$ and CO$_2$; see Figure \ref{fig:gpretrievals}, right panel) are located for temperate, rocky worlds like TRAPPIST-1~e. We do showcase, however, that using data-driven methodologies such as GPs can aid in modeling signals which our stellar models might not be yet ready to account for, allowing us to perform inferences on the transmission spectra which include constraints on the possible atmospheric compositions of TRAPPIST-1~e. A detailed overview of the physical constraints our observations put on possible secondary atmospheric compositions for TRAPPIST-1~e is presented in a companion paper \citep{Glidden2025}; below, we discuss some insights we can extract from our presented observations and methodology, as well as future prospects for further characterization of TRAPPIST-1~e.

\subsection{Stellar contamination beyond 3 um}

One of the most striking features of our observed transmission spectra is their strong epoch-to-epoch wavelength-dependent variations. Intuitively, stellar contamination appearing at wavelengths $<3\ \mu$m at the level observed by prior work \citep{lim:2023, radica:2024} is expected as the result of possible water bands evolving in cold and hot spots in the stellar surface. However, the strong variation at longer wavelengths observed in our transmission spectra along with the inability of stellar models to properly fit the observed variations might seem to counter some of that intuition that would suggest stellar contamination should be smaller at wavelengths past 3 $\mu$m, where water bands might not be strong opacity sources anymore. 

Predictions for stellar contamination for late M-dwarfs such as TRAPPIST-1, however, do point out that depending on the nature of spots the impact at longer wavelengths might not be negligible \citep[see, e.g.,][and references therein]{rackham:2018}. In addition, recent work on modeling the emergent flux of hot and cold spots has also highlighted the need to incorporate the impact of magnetohydrodynamic effects to properly model it \citep[][]{Witzke_2022, norris:2023}. Another line of evidence for expecting variability and stellar contamination at longer wavelengths might also come from recent \textit{JWST} variability studies of brown dwarfs. While colder objects than TRAPPIST-1, the recent variability monitoring over the same wavelength range as our TRAPPIST-1 observations of the brown dwarfs WISE 1049AB and SIMP 0136+0933 reveals ample variability as well in timescales of hours at wavelengths $>3\ \mu$m mainly driven by CH$_4$ and CO variability on those objects \citep{weather1, weather2}. While there is no one-size-fits-all explanation as to the nature of this brown dwarf variability, this is thought to be driven by a complex mixture of various non-equilibrium processes, including condensation and vertical mixing, as well as sampling emergent flux that might be a mixture of contributions from different pressure levels in their atmospheres. It is not unthinkable for TRAPPIST-1 to be developing a similar level of complex physical processes on its surface, which might complicate its modeling even further. 

As demonstrated in our work, methodologies such as GPs exist to marginalize over ``unknown" time-variable signals, which in this work \citep[and in our companion paper;][]{Glidden2025} allowed us to perform various inferences on the transmission spectrum of TRAPPIST-1~e --- including constraints on its possible atmospheric make-up. This methodology, however, has its own limitations. Implicit in our modelling framework introduced in Sections \ref{sec:results} and Appendix \ref{sec:gp-retrieval-framework}, for instance, is the idea that any time-varying signal in the transmission spectrum of TRAPPIST-1~e comes from the star, while any static signal across visits comes from the exoplanetary atmosphere. TRAPPIST-1, however, might possess persistent heterogeneities observable in all the visits, which might also be biasing our exoplanet atmospheric inferences on TRAPPIST-1~e. This is one of the limitations of the presented framework, and one we leave to incorporate in future work. 

Techniques such as those that suggest, e.g., to use TRAPPIST-1~b as a proxy for stellar contamination at all wavelengths and then use that to ``decontaminate" the transmission spectrum of other planets such as that of TRAPPIST-1~e are excellent complementary techniques that could help remove contamination including both, time-varying and persistent components, in a ``model independent" way \citep{roadmap, rathcke:2025}. The observations of \textit{JWST} GO~6456 and 9256  \citep{2024jwst.prop.6456A}, which attempts to use this technique over 15 transits of TRAPPIST-1~b and e will be a perfect dataset to test the GP retrieval methodology introduced in this work, as persistent features should be present in both transmission spectra --- even if time-variable ones vary between close transits of TRAPPIST-1~e and TRAPPIST-1~b. Given the large number of observations, this might also be a perfect program to study possible physical variability mechanisms that might explain the level and wavelength variability on TRAPPIST-1 as well.

\subsection{Primary Atmosphere Constraints on TRAPPIST-1~e}

While with our current data we are unable to distinguish whether TRAPPIST-1~e has an atmosphere or not, our precise 4 visit \textit{JWST}/NIRSpec PRISM spectra, combined with our GP methodology, as introduced in Section \ref{sec:results} allows us to put novel constraints on possible compositions for TRAPPIST-1~e if it were to have an atmospheres. 

As showcased in Section \ref{sec:results}, our work puts in particular strict limits on possible primary, H$_2$-dominated atmospheres present in TRAPPIST-1~e. Prior works attempting to constrain its atmospheric composition were only able to rule out cloud-free, H$_2$-dominated atmospheres \citep[see, e.g.,][]{deWit2018}. Cloudy H$_2$-dominated atmospheres were still allowed, however, as the wavelength range of HST/WFC3 was unable to constrain the amplitude of CH$_4$ and CO$_2$ features (located mostly at wavelengths $>2\ \mu$m) which in such scenario would be very large. On top of this, given stellar activity, studies such as the one from \cite{zhang:2018} suggested the transmission spectra might not be as constraining for TRAPPIST-1~e's exoplanet atmosphere as previously thought --- being, in turn, well explained instead by arising fully from stellar contamination. Applying our GP retrieval methodology to the HST/WFC3 data presented in \cite{zhang:2018}, we indeed reach the same conclusions as those previous works: at cloud-top pressures above about $\sim$1 bar, H$_2$-dominated scenarios are all fully consistent with the data at the 1-2$\sigma$ level.

Using our 4 \textit{JWST} NIRSpec/PRISM visits together with our GP retrieval methodology, however, we were able to showcase that it is very likely TRAPPIST-1~e \textit{does not} possess an H$_2$-dominated atmosphere even in cloudy scenarios, and even in the face of stellar contamination. As we show in Section \ref{sec:results}, the probability that TRAPPIST-1~e has an H$_2$ volume mixing ratio larger than 50\% is less than 1\% given our data and modelling framework. This allows to rule out mixing ratios larger than about 80\% at more than the 3$\sigma$ level. These results are, in turn, in agreement with predictions from \citet{Hori2020DoAtmospheres} that suggest this hydrogen-dominated scenario to be unlikely from hydrodynamic escape calculations. Interestingly, the highest probability regions of the H$_2$ mixing ratios inferred from our \textit{JWST} retrievals (Figure \ref{fig:h2}, right panel) are consistent with those found or predicted to be on Earth, Mars and Venus --- i.e., of order $10^{-6}-10^{-9}$ by volume \citep{ehhalt:1977, patterson:2020, kleinbohl:2024, wang:2025}. A detailed study of the constraints our observations imply for possible secondary atmospheres such as the ones in those planets as traced by other molecules (such as, e.g., CO$_2$ and CH$_4$), including constraints on mean molecular weights are presented in a companion paper \citep{Glidden2025}.

\section{Conclusion} \label{sec:conclusions}

In this work, we present 4 \textit{JWST}/NIRSpec PRISM transmission spectra of TRAPPIST-1~e obtained from mid to late 2023. We show that these transmission spectra, rather than being featureless, exhibit significant variability in both time and wavelength. We interpret this variability as arising from stellar heterogeneities in the host star, TRAPPIST-1 --- i.e., due to the Transit Light Source Effect \citep[TLSE;][]{rackham:2018}. While we can qualitatively explain the observed features in those spectra as arising from possible hot and cold spots on the stellar photosphere, we are unable to fit the observed spectroscopic variations with stellar model atmospheres alone. In order to perform inferences on our transmission spectra and to put constraints on the possible atmospheric make-up of TRAPPIST-1~e, we resort to using Gaussian Processes (GPs) to model the stellar contamination in our spectra, which allows us to perform joint exoplanet atmospheric retrievals on our data and put new constraints on the possible atmospheric compositions of TRAPPIST-1~e. 

We show that with the current dataset we are unable to distinguish between an atmosphere and atmosphereless scenario for TRAPPIST-1~e, despite being able to constrain atmospheric features down to $\sim 50$ ppm at $R=30$ in the 0.6-5 $\mu$m range. This level of precision does allow us, however, to rule out possible cloudy H$_2$, with which we conclude primary atmospheres are unlikely in TRAPPIST-1~e. A detailed study of the possible secondary atmospheres on TRAPPIST-1~e and how they compare to our own Solar System objects are presented in a companion paper \citep{Glidden2025}. We note how the observations of \textit{JWST} GO~6456 and 9256  \citep{2024jwst.prop.6456A} will be critical to constrain both, stellar contamination using methodologies such as the one introduced in this work and others introduced in the literature \citep{roadmap, rathcke:2025}, and the possible atmospheric make-up of TRAPPIST-1~e. Our work does highlight, however, how \textit{JWST} is breaking ground in the study of rocky, habitable-zone exoplanet atmospheric compositions.

\begin{acknowledgments}

We thank the anonymous referee for their helpful and timely comments, which improved this manuscript. {Some/all of the data presented in this article were obtained from the Mikulski Archive for Space Telescopes (MAST) at the Space Telescope Science Institute. The specific observations analyzed can be accessed via \dataset[doi: 10.17909/yzwd-vq54]{https://doi.org/10.17909/yzwd-vq54}. All figures in this paper, along with the associated data, can be accessed at \dataset[doi: 10.5281/zenodo.16125662]{https://doi.org/10.5281/zenodo.16125662}.}\\

This paper reports work carried out in the context of the \textit{JWST} Telescope Scientist Team (\url{https://www.stsci.edu/~marel/jwsttelsciteam.html}, PI: M. Mountain). Funding is provided to the team by NASA through grant 80NSSC20K0586. This work is based on observations made with the NASA/ESA/CSA James Webb Space Telescope. The data were obtained from the Mikulski Archive for Space Telescopes at the Space Telescope Science Institute, which is operated by the Association of Universities for Research in Astronomy, Inc., under NASA contract NAS 5-03127 for \textit{JWST}. These observations are associated with program \#1331 (PI: Lewis). N.H.A. acknowledges support by the National Science Foundation Graduate Research Fellowship under Grant No. DGE1746891. This material is based upon work performed as part of the CHAMPs (Consortium on Habitability and Atmospheres of M-dwarf Planets) team, supported by the National Aeronautics and Space Administration (NASA) under Grant No. 80NSSC23K1399 issued through the Interdisciplinary Consortia for Astrobiology Research (ICAR) program. CIC acknowledges support by NASA Headquarters through an appointment to the NASA Postdoctoral Program at the Goddard Space Flight Center, administered by ORAU through a contract with NASA. DRL acknowledges support from NASA under award number 80GSFC24M0006.


\end{acknowledgments}

%

\vspace{5mm}
\facilities{JWST(NIRSpec)}


\software{JWST Calibration Pipeline \citep{jwstcal}; Eureka \citep{eureka}; transitspectroscopy \citep{transitspectroscopy}, ExoTiC-LD \citep{Grant2024}, ExoTiC-JEDI \citep{exoticjedi}}



\appendix

\renewcommand{\thefigure}{A\arabic{figure}}
\setcounter{figure}{0}
\section{JWST data reduction}\label{sec:red}

\begin{figure*}
\includegraphics[width=\columnwidth]{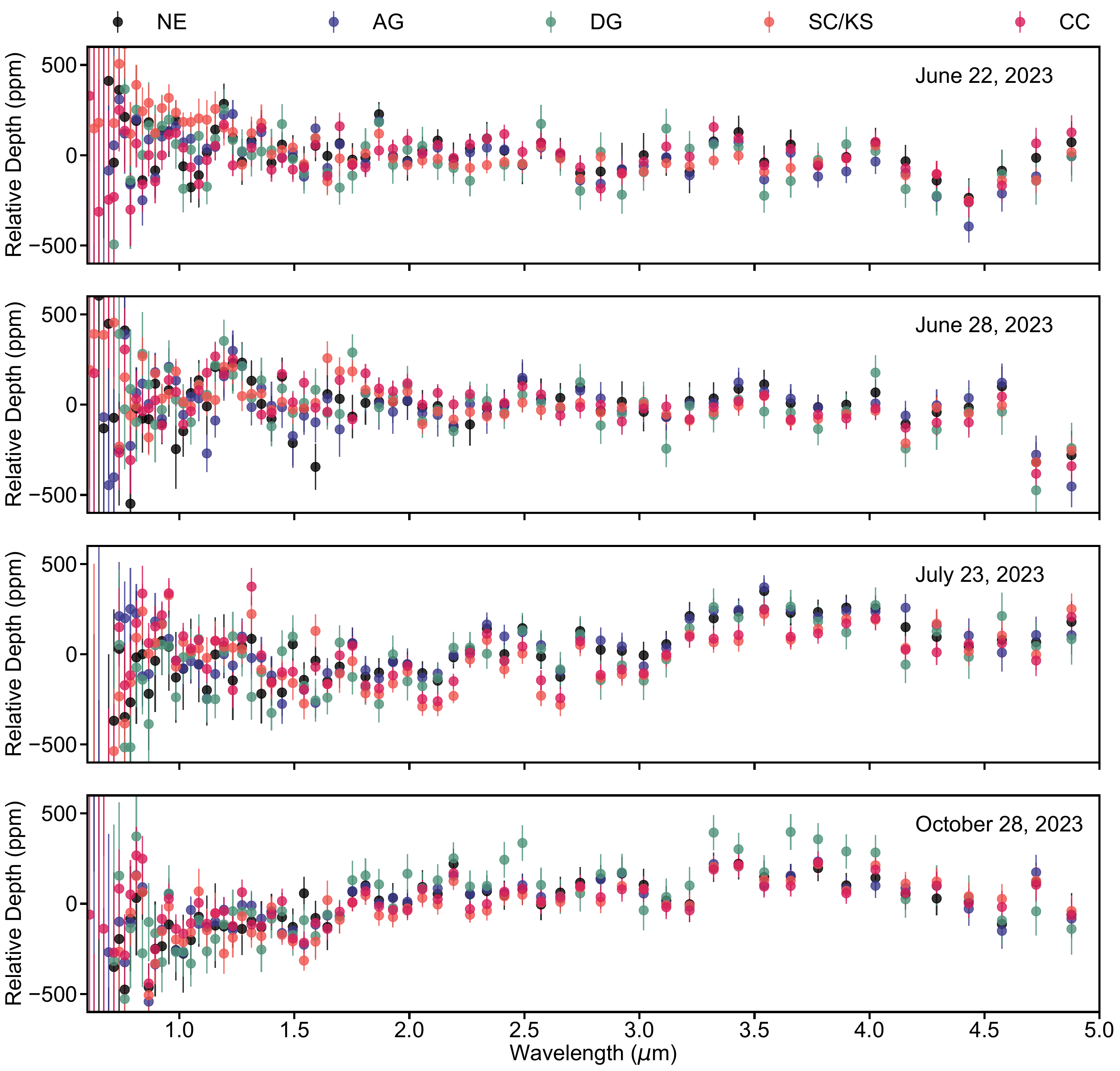}
\caption{\textbf{TRAPPIST-1~e NIRSpec/PRISM transmission spectra on different epochs for different pipelines.} Analogous to Figure \ref{fig:tspectra}, but comparing the transmission spectra obtained by different {pipelines/reductions} in our team. {Note that reductions with the same pipelines by different individuals are plotted with similar colors (i.e., \texttt{transitspectroscopy} reductions by NE and AG are in blue, \texttt{Eureka!} reductions by SC/KS and CC are in orange/red, \texttt{ExoTiC} reduction in green). This is done on purpose for ease of comparison between varying parameters within a given pipeline and comparing reductions with truly different pipelines}. Note we compare the median subtracted transmission spectrum across different pipelines, as our inference described in Section \ref{sec:results} does not depend on the absolute transit depth level but on the \textit{shape} of the transmission spectra. 
\label{fig:tspectra-all}}
\end{figure*}

In what follows, we present all the data reduction steps and pipelines used to obtain the transmission spectrum for TRAPPIST-1~e for our 4 transit observations. In total, we compared our data reduction procedures across 5 different data reduction pipelines. A summary with all the median-subtracted transmission spectra is presented in Figure \ref{fig:tspectra-all}, where we showcase how all our data reduction procedures obtain very similar transmission spectra on our different visits.

We identify each data reduction first by the name of the main pipeline being used, followed by the initials of the co-author(s) performing the reduction. {NE and AG use the \texttt{transitspectroscopy} pipeline. SC/KS and CC use the \texttt{Eureka!} pipeline. DG uses the \texttt{ExoTiC} pipeline. We quantified the agreement between the pipelines by subtracting the resulting relative transmission spectra of NE to the ones produced by SC/KS and DG, and calculating the p-value of a chi-square test on this difference. For all of them we find p-values $>0.4$ --- which is quantitative evidence that the spectra are in agreement with each other.} 

In all the descriptions below, columns refer to pixels that follow the wavelength direction of the spectra, rows are pixels in the cross-dispersion direction.

\subsection{\texttt{transitspectroscopy}, NE reduction} \label{ne-reduction}
The data were reduced using the \texttt{transitspectroscopy} pipeline version 0.4.1 \citep{transitspectroscopy}, which in turn makes use of the \texttt{JWST Calibration Pipeline} \citep{jwstcal}. In particular, for the analysis presented in this work the \texttt{JWST Calibration Pipeline} version 1.12.5 was used. The \texttt{transitspectroscopy} pipeline starts from the uncalibrated data products for each observation (\texttt{*uncal.fits} files), and obtains the rates for each integration using the \texttt{JWST Calibration Pipeline}'s Stage 1 with five modifications: (1) we skip the dark current correction, as the dark current counts are negligible for our short-group-integrations, (2) instead of using the default saturation reference file in the pipeline, we use a modified one that sets the saturation level at 90\% of the level in those files, as this is where we see evident deviations in the ramps from linear even after non-linearity corrections{; the main impact of this change is less outliers are observed in the high resolution transmission spectrum of TRAPPIST-1~e, which translates in better errorbars in particular at wavelengths between 1-2 $\mu$m}, (3) after the superbias step, instead of running the reference pixel step (as NIRSpec/PRISM doesn't have reference pixels) for each group we take the median of the left-most 25 columns and the right-most 25 columns of the spectra (which contain negligible counts from the stellar spectrum) and we remove this value from the entire group to remove group-to-group pedestal changes, (4) background is removed on a group-to-group basis by taking the median at each column of the 2 top pixels and the bottom 2 pixels and removing that value from each column, (5) we perform our own jump-detection for each group as follows. First, we calculate the difference between the fluence $F_{i, g}$ at group $g$ and the corresponding one at group $g+1$ for all integrations $i$, i.e., $D_{g,i} = F_{i, g+1} - F_{i, g}$. Then, a median-filter $M(W, i)$ with a window of $W=200$ integrations is subtracted to this difference time-series, $\tilde{D}_{g,i} = D_{g,i} - M(W, i)$. The median-absolute-deviation-based standard deviation $\sigma$ of this time-series $\tilde{D}_{g,i}$ is calculated, and any values deviating byc more than $10 \sigma$ from it are flagged as jumps in a given group. This process is repeated for $g = 1, 2, 3, 4$. After this procedure, the rates per integration are obtained using the ramp-fitting algorithm in the \texttt{JWST Calibration Pipeline}. To extract the spectrum from those rates per integration, first the spectrum is traced on each integration by finding centroids via a cross-correlation of the profile at each column with a Gaussian, which is then fitted with a B-spline using 8 equally-spaced knots between pixel columns 51 and 491. The median of all of those traces is used as \textit{the} trace for all integrations. Using this, background is removed on each column by removing the median of all pixels more than 10 pixels away from the center of the trace, and 1/f noise is removed following the procedure outlined in \cite{radica:2023}. Finally, the spectrum for each integration is extracted via simple extraction using an aperture with a radius of 7 pixels from the center of the trace.

The transmission spectrum of the planet for each visit is obtained by fitting the wavelength-dependent transit lightcurves. Only the portions within 1-hour from mid-transit of the lightcurves are fitted, and these are, in turn, binned from the original 1.38-second cadence to 13.8-second cadence lightcurves by binning them by a factor of 10 in time. The light curve fits are performed using the \texttt{juliet} library \citep{juliet}. This fits a \texttt{batman} transit lightcurve model \citep{batman} at each wavelength where all parameters are fixed to the ones found in the white-light light curve analysis described in Section \ref{sec:white-light}, except for the planet-to-star radius ratio. Quadratic limb-darkening coefficients are obtained for TRAPPIST-1~e using a 2,566 K, $\log (g)=5.24$ and solar-metallicity PHOENIX model via the \texttt{limb-darkening} library, and through the MC-SPAM algorithm as discussed in \cite{ld}. To account for systematic effects, a GP using the \texttt{celerite} \citep{celerite} library is fitted to each visit and wavelength --- a Mat\`ern 3/2 kernel in time with a time-scale fixed to that found in the white-light lightcurve analysis for each visit described in Section \ref{sec:white-light} is used, with the amplitude of the GP being fitted to each wavelength. A jitter term is fitted and added in quadrature to the photometric errors estimated from the pipeline. The sampling is performed with Dynamic Nested Sampling using the \texttt{dynesty} package \citep{dynesty}.

\subsection{\texttt{transitspectroscopy}, AG reduction} \label{ag-reduction}

A second \texttt{transitspectroscopy} reduction was performed by AG. The steps followed similar ones to the NE reduction calibrating the uncalibrated data products, with the difference that the default saturation reference file was used. The tracing and spectral extraction were the same as the NE reduction, except for the fact that an extraction aperture of 10 pixels was used. The lightcurve fitting setup was the same as the one used for the NE reduction, with the key differences being that (1) no binning in time is performed on the lightcurves and (2) the GP amplitude is fixed to that one found on white-light light curve fits, and the time-scale is fitted at each wavelength. {The motivation for the latter is to test the inverse methodology used by the NE reduction using the same pipeline (which fixes the timescale and fits for the GP amplitude). As it is shown above, both spectra are nearly identical, showcasing that this assumption is not a particularly important one when it comes to retrieven the transmission spectrum.}

\subsection{\texttt{Eureka!}, SC/KS reduction} \label{sc-ks-reduction}

The data were also reduced using the \texttt{Eureka!} pipeline \citep{eureka} version 0.9, which also makes use of the \texttt{JWST Calibration Pipeline}. This particular reduction followed Stage 1 similarly to the \texttt{JWST Calibration Pipeline}, with two main differences: (1) the jump-step is skipped and (2) group-level background subtraction is performed prior to ramp-fitting, using pixels at the top and bottom of each column to remove both background counts and 1/f noise. Then, the standard ramp-fitting algorithm from the \texttt{JWST Calibration Pipeline} is used to obtain the rates per integration. To trace the spectra, a Gaussian was fitted to each column and its parameter was used to obtain the center of the trace at each column. This was done for pixel columns 26 to 451. The spectrum was extracted via optimal extraction using a 3-pixel radius distance from the center of the trace.

The transmission spectrum of the planet was obtained by first binning flux in wavelength as to extract 46 wavelength channels (i.e., by adding 9 pixels in the wavelength direction on each bin). Then, light curves were fitted with the \cite{eureka} utilities, which include a transit lightcurve model, fixing all orbital parameters and limb-darkening coefficients to the same ones used in the previously described reductions, but leaving the planet-to-star radius ratio be a free parameter in the fits. {In addition, a slope and intercept in time are added to the June visits, while a quadratic term was added for the visits of July and October. A} multiplier {was} applied to the error bars of each light curve. In the transit of July 23, where a flare was observed close to mid-transit, the flare event was masked out of the fit {(i.e., the lightcurve was masked from about -0.4 to about -0.1 in Figure \ref{fig:white-light})}.

\subsection{\texttt{Eureka!}, CC reduction} \label{cc-reduction}

A second \texttt{Eureka!} reduction was performed by CC following a similar setup as the SC/KS reduction. On the processing of uncalibrated files to rates per integration, the only difference in this reduction was that a custom bias (implemented in the \texttt{Eureka!} library) was used instead of the \texttt{JWST Calibration Pipeline} standard superbias frame. For spectral tracing, the spectra were traced from pixel column 51 to 451. For the spectral extraction, optimal extraction was also used, but a 4-pixel radius from the traces extraction aperture was used, with background subtraction being performed using all pixels with a radius larger than 8 pixel on each column.

For the lightcurve fitting, the setup was the same as that of SC/KS, with the difference that a quadratic term was used to account for systematic trends and that the lightcurves were fitted at pixel level resolution.

\subsection{\texttt{ExoTiC}, DG reduction} \label{dg-reduction}
The data were also reduced using the \texttt{ExoTiC} reduction framework, with the \texttt{ExoTiC-JEDI} package \citep{exoticjedi} used to go from \texttt{uncal.fits} to 2D images, and a modified version of \texttt{ExoTiC-MIRI} \citep{exoticmiri} used for spectral extraction and light curve fitting. The \texttt{ExoTiC} framework makes use of the \texttt{JWST Calibration Pipeline} version 1.8.2, where the dark current correction was skipped and 1/f subtraction was applied on the group-level using a custom routine detailed in \citet{Alderson2023ERS}. The jump step was used with a threshold of 15-$\sigma$ and the standard ramp-fitting algorithm of the \texttt{JWST Calibration Pipeline} was used. This reduction did not perform tracing on the rates per integration products, and performed aperture extraction using row pixel number 15 as the center, adding all pixel values with a total width of 11-pixels centered on this row. The wavelength dependent light curves were then binned on 0.1$\mu$m bins, and fitted with a methodology similar to that of the \texttt{transitspectroscopy} reduction, with the difference being that instead of fitting a jitter term, a beta-factor was used representing a multiplicative factor on the uncertainties to account for excess white and red noise. All integrations were used to fit the light curves. For each light curve limb-darkening coefficients were calculated using the \texttt{ExoTiC-LD} package \citep{Grant2024} using PHOENIX stellar models \citep{husser2013new} and applying the non-linear 4-parameter law fixing all the coefficients to the model computed values.

\section{Modelling stellar heterogeneities with stellar models} \label{sec:stellarcontamination}

We test the use of both PHOENIX and BT-SETTL stellar models to perform inferences on the transmission spectrum using the NE reduction described above using \textit{exoretrievals} \citep{Espinoza:2019}, though overall we find that the BT-SETTL models are a better fit to the data. {We chose not to use the SPHINX stellar models \citep{Iyer_2023} due to their low resolution (R=250, lower than parts of the PRISM spectrum) and smaller temperature range (2000-4000 K). We fix log g = 5.2396 \citep{agol-t1}, interpolating between the log g = 5 and 5.5 stellar models to get the appropriate model. }We consider hot spots 2750-5000K, and cold spots 2300-2450K for PHOENIX models (which is as low as the PHOENIX model grids reach) and 1500-2450K for BT-SETTL models, against a photospheric temperature of $2566\pm26$K \citep{agol-t1}. We assume that all variations in transit depth with wavelength are due to stellar active regions, modeling the underlying planetary signal as a flat line. However, so that we are not potentially affected by the presence of underlying atmospheric signals, we perform our retrievals with two masks: 1) a 4-4.6 $\mu$m (CO$_2$) mask, and 2) an above 3 $\mu$m (multiple potential species) mask, to both test their consistency and their ability to predict/match the stellar contamination in the masked wavelengths. Interestingly, this second mask acts as a test of what predictive power modeling wavelengths similar to those modeled in the works of \cite{lim:2023} and \cite{radica:2024} for TRAPPIST-1b and c, respectively, with NIRISS/SOSS has on longer wavelengths.

\begin{figure*}
\includegraphics[width=\textwidth]{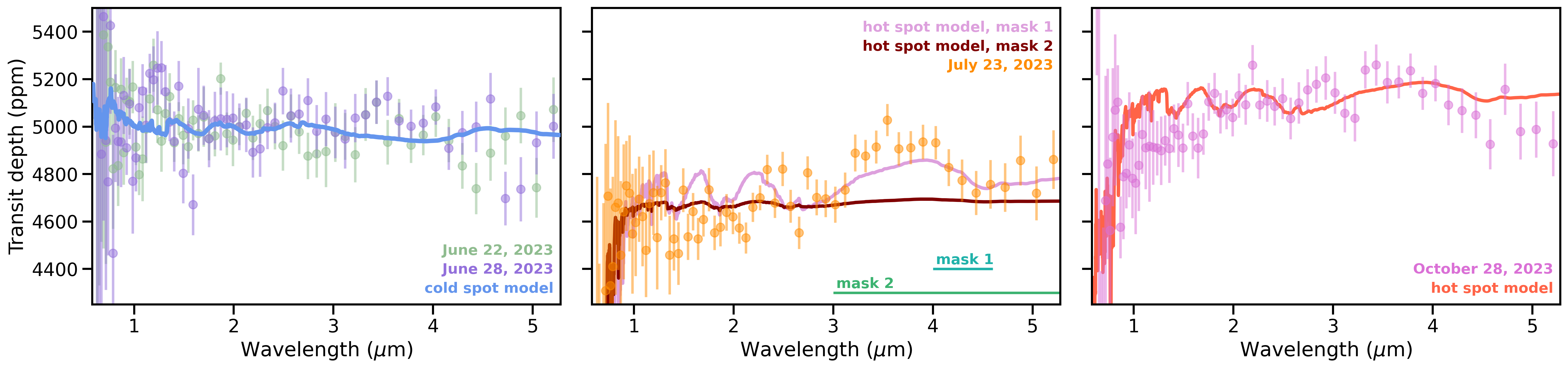}
\caption{\textbf{TRAPPIST-1~e stellar contamination retrievals.} Retrievals are carried out with \textit{exoretrievals} on the NE reduction. All models shown use the BT-SETTL stellar models. Left panel: the first and second visits and the best-fit cold spot model, which is approximately consistent between visits. Middle panel: the third visit, along with hot spot models from both the mask 1 and mask 2 retrieval tests as described in the text. Here it can be seen that a) stellar contamination predicted in the longer wavelengths from that seen in the shorter wavelengths does a poor job at matching the observations, and b) stellar contamination models do a poor job of matching the 1-3 $\mu$m and 3-5 $\mu$m regions simultaneously. Right panel: the fourth visit with the best-fit hot spot model. Similar to point b) from the middle panel, the stellar contamination model cannot fit both the middle wavelengths and the longer wavelengths well.
\label{fig:sc}}
\end{figure*}

We find that all four transmission spectra are consistent with the presence of significant stellar contamination (i.e., the bayesian evidence for all our fits strongly prefers a stellar contamination over a flat line model for our observations). The first two visits (22 and 28 June 2023) are best fit by a small covering fraction ($\sim5$\%) of cold spots with all masks, with the strongest stellar contamination signal in the shortest wavelengths, though the effect is quite small throughout the wavelength range. The covering fractions and spot temperatures between these two visits are consistent with each other, which is interesting given these two visits are separated by close to two rotation periods of TRAPPIST-1 \citep[P$_{\textrm{rot}}=3.3$ days,][]{Luger_2017,Dmitrienko_2018,Brady_2023}, so we are observing the same portion of the stellar surface in both those observations. Importantly, the best-fit temperature for the BT-SETTL model retrievals, which is the model preferred by the data according to the Bayesian evidence, is below the temperature limit of the PHOENIX grid (2100-2200K), which shows that for cold M dwarfs, these generic stellar grids may not cover the entire necessary parameter range to model observed features at least in transmission spectra contaminated by stellar heterogeneities. The first and second visits are shown with the BT-SETTL model for the first visit (whose shape is consistent with that for the second visit) in Figure \ref{fig:sc}, left panel. {The retrieved contamination spectrum and characteristics are largely unchanged between the mask 1 and mask 2 tests.}

The third visit, which has the flare event visible during the transit in Figure \ref{fig:white-light}, and the fourth visit are instead consistent with a surface dominated by hot spots. Both stellar model grids and visits are best-fit by hot spot temperatures of around 2900K for mask 1, though the third visit is consistent with a higher covering fraction than the fourth (11\% vs. 6\%). The fourth visit also agrees with this result for mask 2, but the third visit varies significantly for mask 2. Rather than the parameters found above, fitting only the shorter wavelengths for the third visit instead prefers a very small spot covering fraction with a much higher temperature (1\% covering of 4100K spots), which does a very poor job of matching the contamination seen in the longer, masked wavelengths. This shows us that predicting stellar contamination in the longer wavelengths from their appearance in the shorter wavelengths is not a reliable method. Generally, across these two hot spot dominated visits, we find that our models are a poor fit to the observational data. In order to fit the contamination in the longest wavelengths, the signal in the middle wavelengths is significantly overestimated, or the inverse, such that we cannot fit the full wavelength range well with our models. We are confident that this mismatch between the models and features in the longest wavelengths is \textit{not} due to atmospheric features, since these features are only visible in the case of a hot spot dominated stellar surface. We show the third visit, along with the best-fit BT-SETTL hot spot models for mask 1 and mask 2, and the fourth visit with the best-fit BT-SETTL hot spot model, in Figure \ref{fig:sc}, middle and right panel respectively.

We conclude that, although our data presents strong evidence for stellar contamination in the transmission spectrum of TRAPPIST-1~e, performing inferences in the transmission spectrum is not straightforward with our current stellar models for TRAPPIST-1. We are perhaps seeing the limits of our current methods, since the use of photospheric models for the modeling of magnetic active regions like spots and faculae is inherently incorrect \citep[see e.g.][]{Witzke_2022}. To get to the point where we are able to correct for stellar contamination to the precision necessary to detect the $\sim$ 10's of ppm signals from terrestrial exoplanet atmospheres, we must work towards creating better stellar active region models, or change our approach in tackling the problem of stellar contamination.

\section{GP Atmospheric Retrieval Framework} \label{sec:gp-retrieval-framework}

We implemented the retrieval methodology based on equation (\ref{eq:log}) as follows. First, we convert our measured transit depths and errors to log-space using the transformations $\log \delta_{v,i}(\lambda_i)$ and $\sigma_{\log \delta_{v,i}} = \sigma_{\delta_{v,i}} / \delta_{v,i}$, with the former being the depths in log-space and the latter being their errors --- obtained through the delta method. The log-likelihood is then computed independently for each visit using \texttt{george}'s \texttt{log\_likelihood} function, which are then added to form the total log-likelihood. We perform our inferences using dynamic nested sampling via the \texttt{dynesty} library \citep{dynesty}. 5,000 live points are set initially, with the \texttt{multi} bound and the random walk (\texttt{rwalk}) sampler.

For our GP we use a Mat\`ern 3/2 kernel, where instead of using the \texttt{celerite} \citep{celerite} package approximation we use the \texttt{george} implementation which handles the exact covariance for this kernel \citep{george}. The hyperparameters of our GP $\theta_{c,v}=\{A_{c,v}, \ell_{c,v}\}$, include a visit-dependent amplitude $A_{c,v}$ with a uniform prior from 0 to 10 dex, and a visit-dependent length-scale $\ell_{c,v}$ (in microns) with a uniform prior of 0 to 100 $\mu$m for the GP. The jitter term per visit has a uniform prior between 0 and 1000 ppm. The offset $C_v$ has a uniform prior as well of -3000 to 3000 ppm.

For the exoplanet atmospheric model $S(\lambda_i)$ we draw forward models from the \texttt{POSEIDON} library \citep{poseidon1, poseidon2} at a resolution of $R=10,000$, which we then degrade at each step of the sampling using the \texttt{POSEIDON.instrument.make\_model\_data} function, with appropiate inputs calculated via the \texttt{POSEIDON.core.init\_instrument} function tailored for NIRSpec/PRISM. Motivated by the work of \cite{Lin:2021} and \cite{LY:2023} we consider H$_2$, CO$_2$, CH$_4$, H$_2$O, N$_2$, O$_2$, O$_3$, N$_2$O and CO as the possible species in our modeling framework. Following \cite{bennekeseager:2012}, we use a centered-log-ratio transformation to model the mixing ratios in order to consider any of the spectrally active molecules in our retrievals to be the background gas. At each step of the sampler, we draw

\begin{equation}
\xi_i = \ln X_i/g(\mathbf{x}) \label{xi-mr}
\end{equation}
for each element $i$ except for H$_2$, which we derive from the constraint that $\sum \xi_i = 0$. With this, we calculate $g(\mathbf{x}) = 1 / \sum \exp \xi_i$, which allows us, using equation (\ref{xi-mr}), to calculate the individual mixing ratios $X_i$ which are feeded to \texttt{POSEIDON} to obtain a forward model. We set uniform priors of -22.47 to 24.17 for the $\xi_i$, however, we reject any samples that give rise to $X_i<10^{-12}$ or $X_i>1$. We also fit for a cloud-top pressure with a log-uniform prior from 10$^{-7}$ to 100 bar, the reference pressure which also has a log-uniform prior from 10$^{-7}$ to 100 bar, and use an isothermal temperature/pressure profile, with the temperature being a free parameter as well with a uniform prior between 100 and 300 K. For the star, we use a radius of 0.11697$R_\odot$, effective temperature of 2559, Fe/H of 0.04 and log-gravity of 5.21. For TRAPPIST-1~e, we use a radius of $0.917985R_\oplus$, mass of $0.6356 M_\oplus$ and equilibrium temperature of 255 K. For retrievals incorporating GPs and exoplanet atmospheric models, thus, the total number of free parameters is 27.

For the stellar contamination modeling $\epsilon_{c,v}(\lambda_i)$, we set uniform priors on the temperature of cold ``spots" from 1500 to 2450 K, and for hot ``spots" from 2750 to 5000 K. Spot covering fractions also have uniform priors between 0 and 1, and we reject samples that lead to sums of hot and cold covering fractions larger than 1. We use BT-SETTL stellar models to model both spots and the stellar photosphere for TRAPPIST-1 \citep{bt-settl}. {We use the same stellar parameters for TRAPPIST-1 as in Appendix \ref{sec:stellarcontamination}}. 

For the flat-line retrievals presented in Section \ref{sec:results}, we set $S(\lambda_i) = \epsilon_{c,v}(\lambda_i) = 0$. We set a uniform prior on the offset $C_v$ of $\delta_{t1e}-3000$ to $\delta_{t1e}+3000$ ppm, with $\delta_{t1e} = 5176.8$ ppm (which is consistent with the planet-to-star radius ratio used to initialize \texttt{POSEIDON}, as discussed above). 

\subsection{JWST NIRSpec/PRISM retrievals posteriors}

\begin{figure*}
\includegraphics[width=\textwidth]{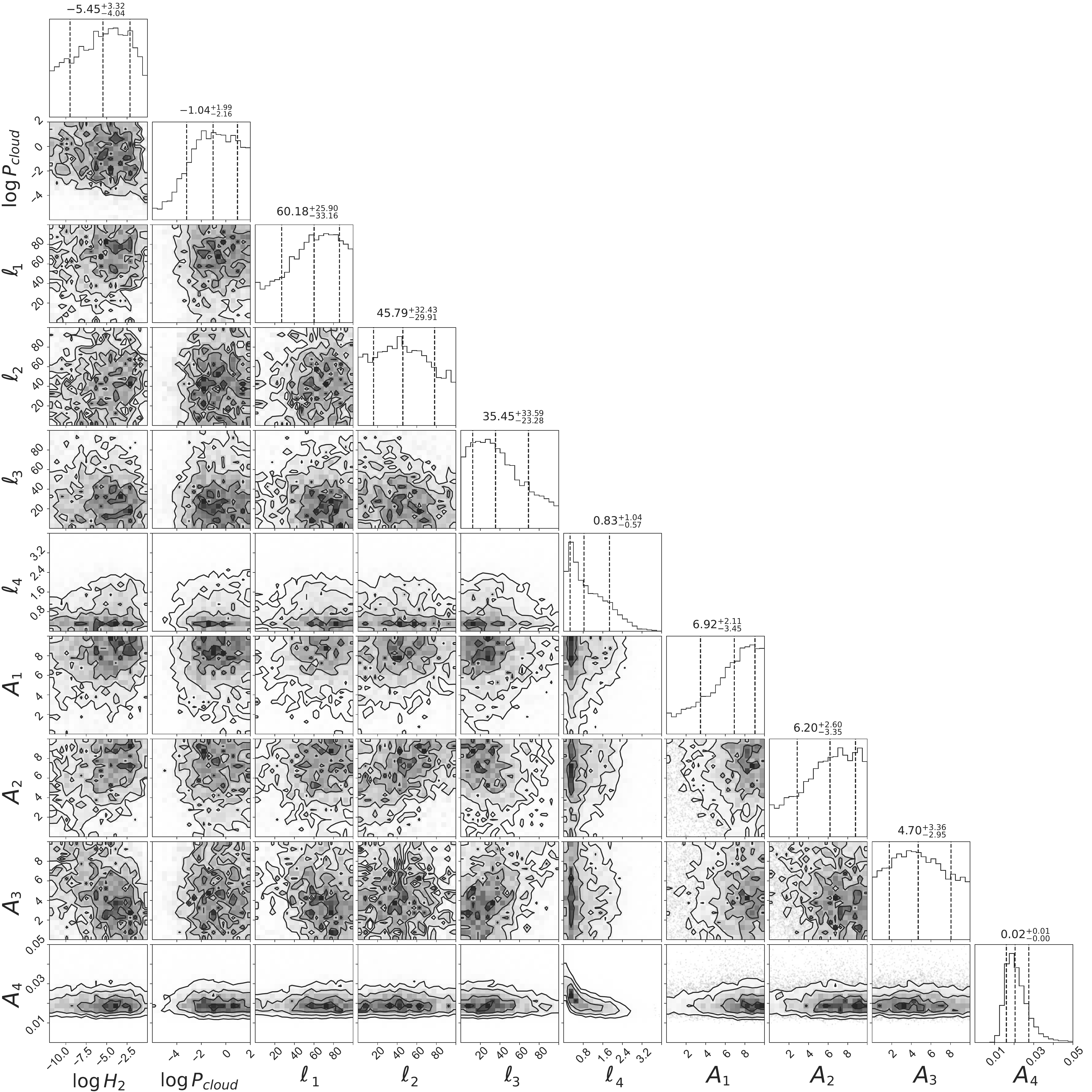}
\caption{{\textbf{Atmospheric and GP hyperparameter posterior distributions.} Posterior distribution of some of our JWST NIRSpec/PRISM retrieval parameters, compared to the atmospheric parameters we constrain in this work in Figure \ref{fig:h2}, H$_2$ and cloud-top pressure/surface pressure. Note how the GP hyperparameters (lengthscales $\ell_i$ and amplitudes $A_i$) don't show strong correlations with the atmospheric parameters.}
\label{fig:cornerplot}}
\end{figure*}

{In Figure \ref{fig:cornerplot} we show a subset of our posterior distributions for the JWST NIRSPec/PRISM atmospheric retrievals following the above mentioned framework, which has several interesting insights. First, note how the atmospheric parameters (cloud-top pressure or surface pressure along with H$_2$ mixing ratio) do not show strong correlation with the GP hyperparameters (lengthscales $\ell_i$ and amplitudes $A_i$). Second, note how the GP lengthscales ($\ell_i$) get progressively smaller for the last two visits --- where we see the most prominent trends in the transmission spectrum when it comes to stellar contamination. For visits 1 and 2, the lengthscales are large so the GP 
is mostly a nearly flat-line --- which is exactly what is observed in Figure \ref{fig:gpretrievals}. For visits 3 and particularly for visit 4, the lengthscale is smaller, which is reflected in the behavior of the GP in the same figure.}

\subsection{HST/WFC3 retrievals}

\begin{figure*}
\includegraphics[width=\textwidth]{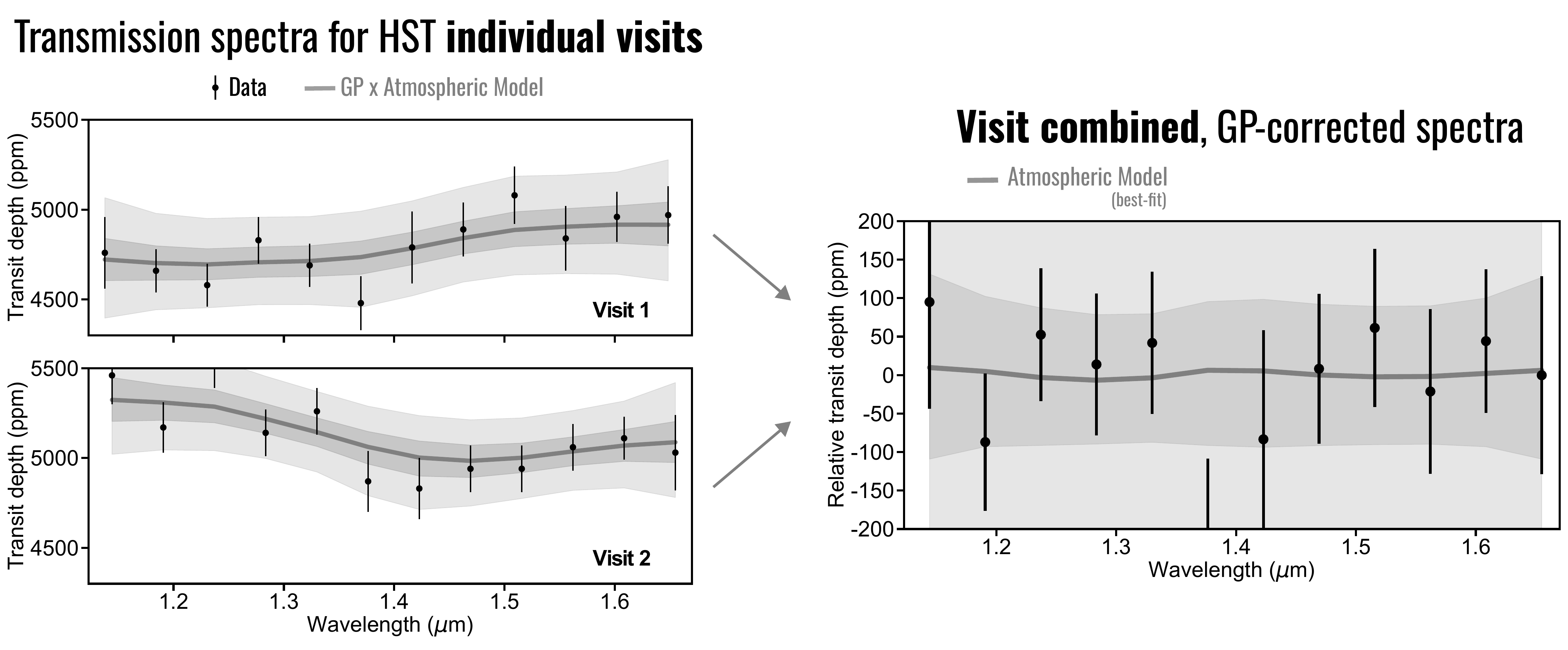}
\caption{\textbf{The TRAPPIST-1~e HST/WFC3 transmission spectra presented in \cite{zhang:2018} interpreted with Gaussian Processes and an atmospheric model.} (\textit{Left}) Transmission spectra of the 2 HST/WFC3 visits (black points with error bars) modeled with a GP times an atmospheric model (grey); analogous to the top panels in Figure \ref{fig:gpretrievals}. Bands represent the 1 and 3-sigma credibility bands. (\textit{Right}) Visit combined transmission spectrum obtained by averaging the two WFC3 visits after correcting for the modeled GP component (black points with error bars). The atmospheric model is in grey. Bands represent the 1 and 3-sigma credibility bands. This is analogous to the bottom panel of Figure \ref{fig:gpretrievals}. Note we use the very same y-limits in this Figure and in Figure \ref{fig:gpretrievals}. 
\label{fig:gpretrievals-hst}}
\end{figure*}

Finally, for the HST/WFC3 retrievals whose posterior distributions are compared to the \textit{JWST} ones Section \ref{sec:results} using the data in \cite{zhang:2018}, we use the very same methodology and priors outlined above for the \textit{JWST} retrievals, although this only incorporates data for the 2 visits analyzed in that work. The retrievals also use models generated at a resolution of $R=10,000$, which are binned to the appropriate HST/WFC3 G141 grism. We present the analogous of Figure \ref{fig:gpretrievals} presented in Section \ref{sec:results} for our \textit{JWST} results but for the retrievals performed on this HST/WFC3 data in Figure \ref{fig:gpretrievals-hst}. 

\bibliography{sample631}{}
\bibliographystyle{aasjournal}



\end{document}